\documentclass[aps,prb,twocolumn,showpacs,superscriptaddress]{revtex4}
\usepackage{bm,color,amsmath,amssymb,graphicx}
\usepackage{hyperref}

\begin{document}

\title{Zoology of Fractional Chern Insulators}

\author{Yang-Le Wu}
\affiliation{Department of Physics, Princeton University, Princeton, NJ 08544}
\author{B. Andrei Bernevig}
\affiliation{Department of Physics, Princeton University, Princeton, NJ 08544}
\author{N. Regnault}
\affiliation{Laboratoire Pierre Aigrain, ENS and CNRS, 24 rue Lhomond, 75005 Paris, France}

\begin{abstract}
We study four different models of Chern insulators in the presence of strong 
electronic repulsion at partial fillings. We observe that all cases 
exhibit a Laughlin-like phase at filling fraction $1/3$. We provide evidence 
of such a strongly correlated topological phase by studying both the energy 
and the entanglement spectra. In order to identify the key ingredients of the 
emergence of Laughlin physics in these systems, we show how they are affected 
when tuning the band structure. We also address the question of the relevance 
of the Berry curvature flatness in this problem. Using three-body interactions, we 
show that some models can also host a topological phase reminiscent of the 
$\nu=1/2$ Pfaffian Moore-Read state. Additionally, we identify the
structures indicating cluster correlations in the entanglement spectra.
\end{abstract}

\maketitle

\section{Introduction}

First introduced by Haldane in his 1988 paper~\cite{Haldane88:Honeycomb}, the 
Chern insulator is the simplest example of a topological insulator. The 
topological insulating phase is characterized by the (non-zero) Chern number 
of the occupied band, and it exhibits a non-zero integer Hall conductance 
$\sigma_{xy}$ at zero magnetic field. The Hall conductance can be attributed 
to the metallic edge states protected by the non-trivial topology of the bulk.
Interest in such states was boosted by the theoretical 
prediction~\cite{Kane05:KM,Kane05:KM2,Bernevig06:QSH,Bernevig06:BHZ} and 
subsequent experimental 
realization~\cite{Konig07:QSH,Hsieh08:BiSb} of time-reversal topological 
insulators. In its simplest form in two dimensions, this type of insulator 
consists of two 
copies (spin up and down) of a Chern insulator. Comprehensive studies have been 
carried out~\cite{Hasan10:RMP,Hasan11:AnnRev}, and a complete classification of 
topological band theories with time-reversal and charge-conjugation symmetries
in all dimensions has been 
established~\cite{Schnyder08:Table,Qi08:Table,Kitaev09:Table}.
More than two decades after its first introduction, however, the theoretical 
study of topological insulators is still mostly limited to the single-particle 
regime, with interaction effects playing a subleading role.

Recent developments have shown that strong electronic interactions 
open up interesting new possibilities. Novel phases resembling the fractional 
quantum Hall (FQH) effect have been identified in lattice models at zero magnetic 
field. Several authors have reported the discovery of the `fractional Chern 
insulator' (FCI)~\cite{Neupert11:FCI,Sheng11:FCI,Regnault11:FCI}. They found a 
FQH phase of interacting electrons on a checkerboard lattice at filling $1/3$ and 
\emph{zero} magnetic field. The feature of this phase is an almost 
3-fold degenerate incompressible ground state with Hall conductance 
$\sigma_{xy}=1/3$, reminiscent of the FQH Laughlin state on a torus. 
A similar state has been found on a triangular lattice as well~\cite{Venderbos11:t2g}.
In parallel, FCI phases of interacting bosons have been 
realized~\cite{Wang11:FCI-Boson,Wang11:MR}, a time-reversal-symmetric fractional 
topological liquid state has been constructed from two copies of 
FCI~\cite{Neupert11:Z2,Santos11:BF}, and an integer quantum Hall effect has 
been found in a half-filled Hubbard model in coexistence with an Ising 
ferromagnetic order~\cite{Neupert11:Hubbard}.

The resemblance of the FCI phase to the conventional FQH effect has 
been justified from several perspectives. First, the FCI phase has 
quasiparticle excitations with fractional 
statistics~\cite{Regnault11:FCI,Bernevig11:Counting}, 
and the quasiparticles are governed by the admissible counting rules first 
found in the study of the FQH effect~\cite{Bernevig08:Jack,Bernevig08:Jack2}.
Second, in the limit of long wavelength 
and uniform Berry curvature, the projected single-particle density operators 
form a closed Lie algebra~\cite{Parameswaran11:W-inf,Murthy11:CF}. This algebra 
has the same structure as the Girvin-MacDonald-Platzman 
algebra of magnetic translations and projected density operators in the
FQH effect~\cite{Girvin86:GMP}. Third, Wannier functions maximally localized in 
one dimension have been explicitly constructed from the single-particle states 
of a Chern band~\cite{Qi11:Wavefunction}.
Finally, various constructions of 
FQH-analogue wave-functions for the ground state of FCI have been proposed, using either Wannier 
functions~\cite{Qi11:Wavefunction} or a parton 
approach~\cite{Lu11:Parton,McGreevy11:Parton,Vaezi11:Parton}.

In this paper we show that the FCI phase is present in the Haldane model on 
the honeycomb lattice~\cite{Haldane88:Honeycomb}, in a two-orbital model that 
resembles half (spin-up) of the Mercury-Telluride two-dimensional topological 
insulator~\cite{Bernevig06:BHZ}, in the Kagome lattice model 
with spin-orbit coupling~\cite{Tang11:Kagome}, and in the spin-polarized ruby 
lattice model~\cite{Hu11:Ruby}. These models allow us to study FCI in different 
physical situations such as different lattices or different number of Bloch bands. 
Working in a single flattened band, we find in each model a more or less 
robust Laughlin FQH phase at $1/3$ filling in the 
presence of repulsive two-body nearest-neighbor interactions. The 3-fold 
degenerate ground states are separated from the excited states by a finite gap 
and flow into each other upon flux insertion with a period of 3 fluxes. This 
signals a Hall conductance $\sigma_{xy}=1/3$. We identify hallmarks of 
fractional excitations of the Laughlin $1/3$ universality class in the energy 
spectrum as well as the entanglement spectrum. We then discuss the stability 
of the topological ground state under parameter variation, and test its 
correlation with the anisotropy of the Berry curvature.
We also show that another FCI phase reminiscent of the Pfaffian 
Moore-Read FQH state~\cite{Moore91:MR} is present in the Kagome and ruby
lattice models at half filling of the valence band. Finally, we highlight 
several structures in the higher levels of the particle entanglement 
spectrum~\cite{Li08:ES,Sterdyniak11:PES} of the ground state at filling $1/3$ 
that may serve as a hint for the stable 
existence of other FQH states at other fillings, such as the 
Read-Rezayi series~\cite{Read99:RR}.

\section{Haldane Model}\label{sec:haldane}
\begin{figure}[]
\centering
\includegraphics[]{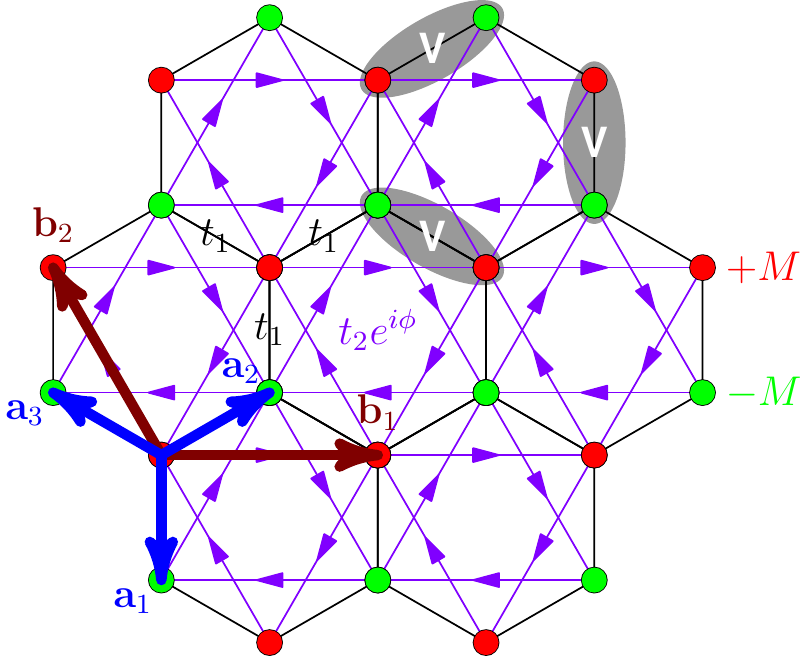}%
\caption{\label{fig:haldane} The Haldane model on the honeycomb lattice. $A$ 
and $B$ sublattices are colored in red and green, respectively. The lattice 
translation vectors are $\mathbf{b}_1=\mathbf{a}_2-\mathbf{a}_3$ and 
$\mathbf{b}_2=\mathbf{a}_3-\mathbf{a}_1$. The hopping between NN is $t_1$, and 
the hopping between NNN is $t_2e^{i\phi}$ in the direction of arrows.
The density-density repulsion between nearest neighbors is depicted in gray 
ellipses.
}
\end{figure}

The Haldane model~\cite{Haldane88:Honeycomb} is the first studied example of a 
topological insulator. We would like to see if this model can host fermionic 
FCI phases (the bosonic version has recently been reported in 
Refs.~\onlinecite{Wang11:FCI-Boson,Wang11:MR}).
We adopt the honeycomb lattice layout from Ref.~\onlinecite{Neupert11:FCI}.
As shown in Fig.~\ref{fig:haldane}, the 
two sublattices $A$ and $B$ are connected by the vectors $\mathbf{a}_1=(0,-1)$, 
$\mathbf{a}_2=(\sqrt{3}/2,1/2)$, $\mathbf{a}_3=(-\sqrt{3}/2,1/2)$. We 
define the lattice translation vectors $\mathbf{b}_1=\mathbf{a}_2-\mathbf{a}_3$,
$\mathbf{b}_2=\mathbf{a}_3-\mathbf{a}_1$. The Haldane 
model~\cite{Haldane88:Honeycomb} has real hopping amplitude $t_1$ between nearest 
neighbors (NN), complex hopping amplitude $t_2e^{\pm i\phi}$ between 
next-nearest neighbors (NNN), and an inversion-breaking sublattice potential $M$.

After a Fourier transform to the first Brillouin zone and a gauge 
transform on the $B$ sublattice $\psi_{\mathbf{k},B}\rightarrow 
e^{i\mathbf{k}\cdot(\mathbf{b}_1+2\mathbf{b}_2)/3}\psi_{\mathbf{k},B}$,
the single-particle Hamiltonian can be put in Bloch form as
$H=\sum_\mathbf{k} (\psi^\dagger_{\mathbf{k},A}, \psi^\dagger_{\mathbf{k},B})
h(\mathbf{k})(\psi_{\mathbf{k},A}, \psi_{\mathbf{k},B})^\text{T}$. 
Here the lattice momentum 
$\mathbf{k}=(\mathbf{k}\cdot \mathbf{b}_1,\mathbf{k}\cdot \mathbf{b}_2)
\equiv (k_x,k_y)$ is summed over the first Brillouin zone, and the $h(\mathbf{k})$ 
matrix can be expressed in terms of the identity and 3 Pauli matrices, 
$h(\mathbf{k})=d_0\mathbb{I}+\sum_i d_i\sigma_i$, where 
\begin{align}
d_0&=2t_2\cos\phi[\cos k_x + \cos k_y + \cos(k_x+k_y)],\\\nonumber
d_x&=t_1[1+\cos(k_x+k_y)+\cos k_y],\\\nonumber
d_y&=t_1[-\sin(k_x+k_y)-\sin k_y],\\\nonumber
d_z&=M+2t_2\sin\phi[\sin k_x + \sin k_y - \sin(k_x+k_y)].
\end{align}

The single-particle Hamiltonian has inversion symmetry at $M=0$ and the 
3-fold point group symmetry of the honeycomb lattice. At $M=0$, inversion 
exchanges the two sublattices and transforms the annihilation operators by
$(\psi_{\mathbf{k},A},\psi_{\mathbf{k},B})^\text{T}\rightarrow 
\sigma_x(\psi_{-\mathbf{k},A},\psi_{-\mathbf{k},B})^\text{T}$.
The 3-fold rotation generates the cyclic permutation of the lattice 
translation vectors $\mathbf{b}_1\rightarrow\mathbf{b}_2\rightarrow
-\mathbf{b}_1-\mathbf{b}_2\rightarrow \mathbf{b}_1$ on each sublattice 
and thus transforms the wave vectors by $(k_x,k_y)\rightarrow (k_y,-k_x-k_y)$. 
Therefore, the Bloch Hamiltonian has the following two symmetries:
\begin{align}
h(k_x,k_y)&=\sigma_xh(-k_x,-k_y)\sigma_x,\\\nonumber
h(k_x,k_y)&=U^\dagger(k_x,k_y)h(k_y,-k_x-k_y)U(k_x,k_y),
\end{align}
where $U(k_x,k_y)$ 
is a diagonal $2\times2$ unitary matrix, with $[1,e^{i(k_x+k_y)}]$ on the 
diagonal. When the system is put on the lattice of finite size $N_x\times N_y$ 
with periodic boundary, the 3-fold rotation symmetry is lifted, unless 
$N_x=N_y$. 

To focus on the effect of interactions without being distracted by 
single-particle dispersion, we always take the flat-band limit of the 
insulator, i.e. replace the original Bloch Hamiltonian 
$h(\mathbf{k})=\sum_n E_n(\mathbf{k})P_n(\mathbf{k})$ by 
$\sum_n E_n(0)P_n(\mathbf{k})$, where $P_n(\mathbf{k})$ is the projector onto 
the $n$-th band. We then send the band gap to infinity and work 
directly in the lowest band, in the same spirit of the lowest Landau level 
projection routinely adopted in the FQH literature. We then add a 
density-density repulsion between nearest neighbors. Since 
a flattened band does not provide an energy scale, we are free to set the 
interaction strength to unity. After the aforementioned gauge transform on 
$\psi_{\mathbf{k},B}$, the interaction term can be written in the sublattice 
basis as
\begin{equation}
\frac{1}{N}\sum_{\{\mathbf{k}_i\}} 
\psi^\dagger_{\mathbf{k}_3A}\psi^\dagger_{\mathbf{k}_4B}
\psi_{\mathbf{k}_2B}\psi_{\mathbf{k}_1A}
\delta_{\mathbf{k}_1+\mathbf{k}_2-\mathbf{k}_3-\mathbf{k}_4}^{\text{mod} 2\pi}
V_{\mathbf{k}_1\mathbf{k}_2\mathbf{k}_3\mathbf{k}_4},
\end{equation}
where
\begin{equation}\label{eq:haldane-interaction-factor}
V_{\mathbf{k}_1\mathbf{k}_2\mathbf{k}_3\mathbf{k}_4}=
1 + e^{i(\mathbf{k}_2-\mathbf{k}_4)\mathbf{b}_2}
+e^{i(\mathbf{k}_2-\mathbf{k}_4)(\mathbf{b}_1+\mathbf{b}_2)},
\end{equation}
as illustrated in Fig.~\ref{fig:haldane}.

\subsection{Ground state at $1/3$ filling}

\begin{figure}[]
\includegraphics{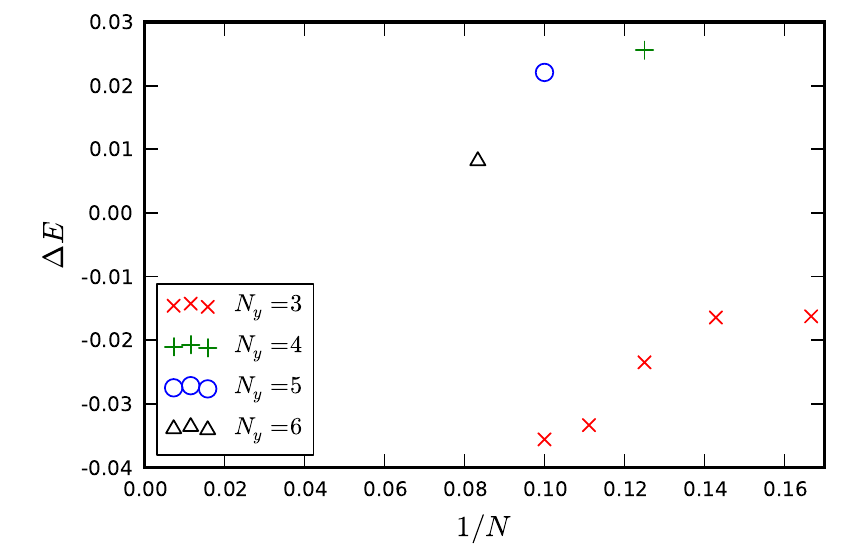}
\caption{\label{fig:haldane-scaling} Energy gap $\Delta E$ for different 
system sizes and aspect ratios of the Haldane model. The gap is defined as the 
energy difference between the first excited state and the highest of the 
3-fold ground states. A more precise definition of $\Delta E$ is given in 
Section~\ref{sec:haldane-paramdep}. In each case, $N_x=3N/N_y$. The negative 
values of $\Delta E$ in the $N_y=3$ group mean that the actual ground state is 
not in the momentum sectors predicted by the theoretical counting rule, i.e. 
the gap above the supposed topological ground state closes.
}
\end{figure}

We diagonalize the interacting Hamiltonian in the 
flattened lowest band at filling $1/3$. We show the energy spectrum of 
$N=8,10,12$ particles on the
$N_x\times N_y=6\times \frac{N}{2}$ lattice in Fig.~\ref{fig:864,1065,1266}. 
The calculations are performed with $(t_1,t_2,M,\phi)=(1,1,0,0.13)$. The 
particular choice of parameters will be 
discussed later. In the three cases ($N=8,10,12$), a 3-fold degenerate ground 
state is seen at total momenta $\{(0,0),(2,0),(4,0)\}$, $\{(1,0),(3,0),(5,0)\}$, 
and $\{(0,0),(0,0),(0,0)\}$, respectively. This agrees perfectly with the 
$(1,3)$-admissible 
counting proposed in Ref.~\onlinecite{Regnault11:FCI} and recently developed
in Ref.~\onlinecite{Bernevig11:Counting}. The finite-size scaling 
of the energy gap $\Delta E$ is shown in Fig.~\ref{fig:haldane-scaling}. 
The energy gap does not seem to remain open in the thermodynamic limit even if 
the aspect ratio $N_x/N_y$ remains finite. 
Nonetheless, as discussed in the following sections, we find solid evidence 
for the topological nature of the ground state; the detailed investigation 
into the energetics will be presented in future work.

\begin{figure}[]
\includegraphics{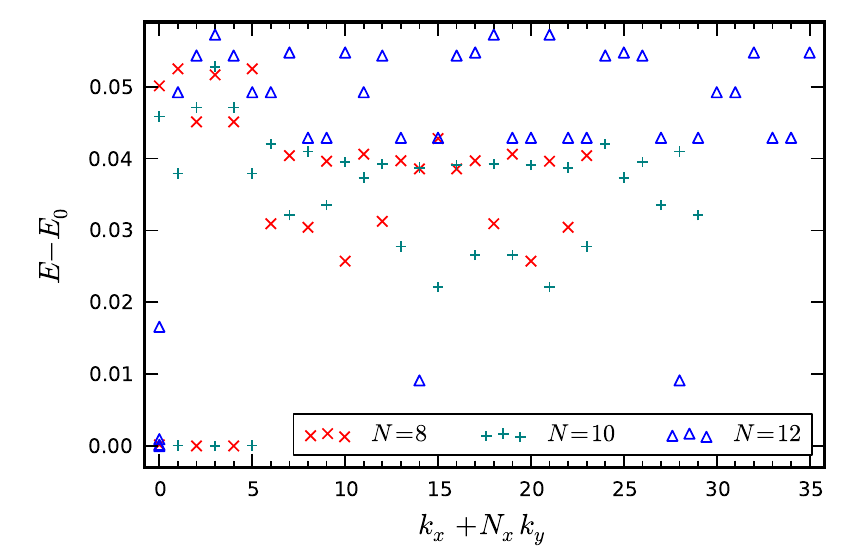}
\caption{\label{fig:864,1065,1266} Low energy spectrum of the Haldane model 
with $N=8$ (marked by red crosses), $N=10$ (green plus signs), and $N=12$ 
(blue triangles) particles on the $N_x\times N_y=6\times \frac{N}{2}$ lattice, 
with energies shifted by $E_0$, the lowest energy for each system size.
We only show the lowest excited level in each momentum sector in addition to 
the 3-fold ground state. 
}
\end{figure}

\begin{figure}[]
\includegraphics{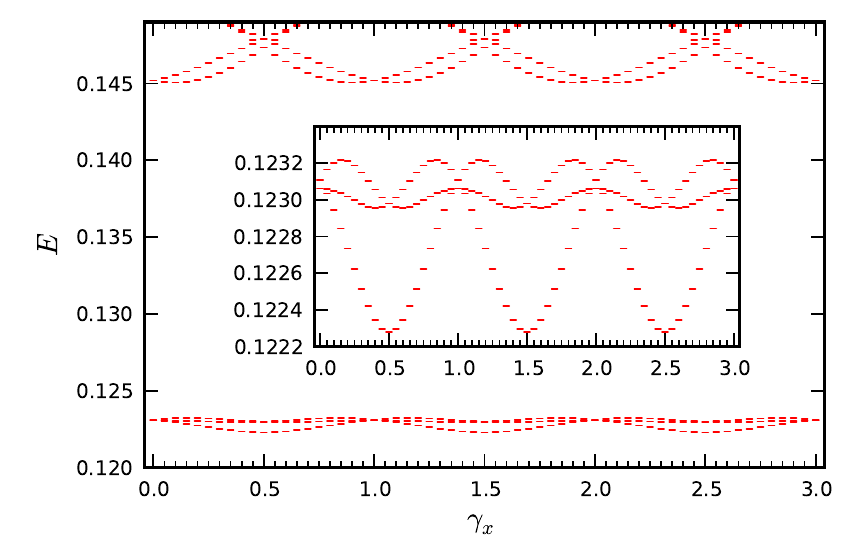}
\caption{\label{fig:1065twisted} Spectral flow of the low-lying
states of the Haldane model with $N=10$ particles on the $N_x\times N_y=6\times 5$ 
lattice upon flux insertion along the $x$ direction. $\gamma_x$ counts the 
number of fluxes inserted. The 3-fold ground states 
flow into each other, and do not mix with higher states during flux insertion. 
It takes 3 full fluxes for the 3-fold states to return to the original 
configuration (inset).}
\end{figure}

The three degenerate ground states exhibit spectral flow upon flux insertion. For 
example, inserting a unit flux in the $x$ 
direction shifts the single particle momentum 
$k_x \rightarrow k_x + 2\pi\gamma_x$, with $\gamma_x$ 
going from 0 to 1; this induces the spectral flow within the 3-fold ground 
state (Fig.~\ref{fig:1065twisted}). Upon insertion of 3 full fluxes, the 3 
degenerate states restore their original configuration. Given the unit
Chern number of the valence band, we conclude the system has Hall conductance 
$\sigma_{xy}=1/3$. 

We have also checked the effect of density-density repulsion between NNN on 
the $1/3$ phase. Overall, this additional interaction term weakens the 
Laughlin-like phase. These results are in agreement with those of Sheng et 
al.~\cite{Sheng11:FCI} for the checkerboard model.

\subsection{Quasihole excitations}

\begin{figure}[]
\includegraphics{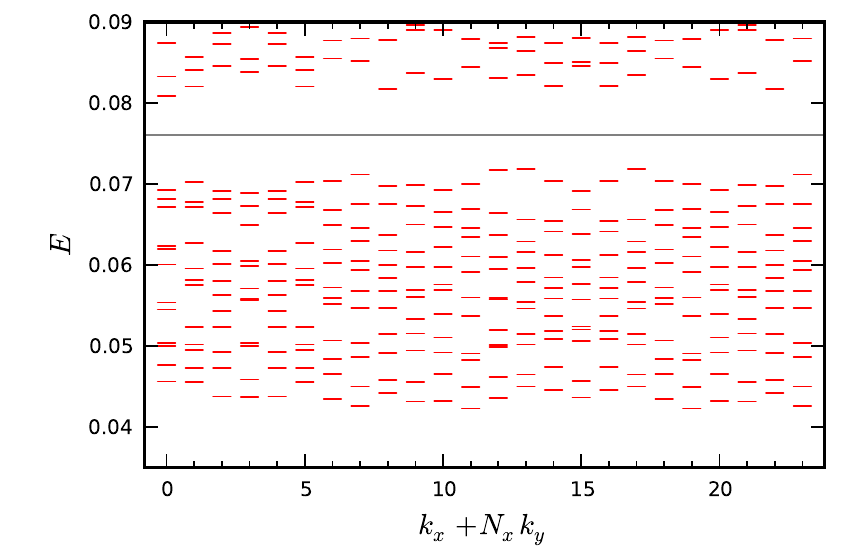}
\caption{\label{fig:764} Low energy spectrum of the Haldane model with $N=7$ 
particles on the
$N_x\times N_y=6\times 4$ lattice. The number of states below the gray line is 
$12$ in each momentum sector, in agreement with the $(1,3)$-admissible 
counting rule.}
\end{figure}

\begin{figure}[]
\includegraphics{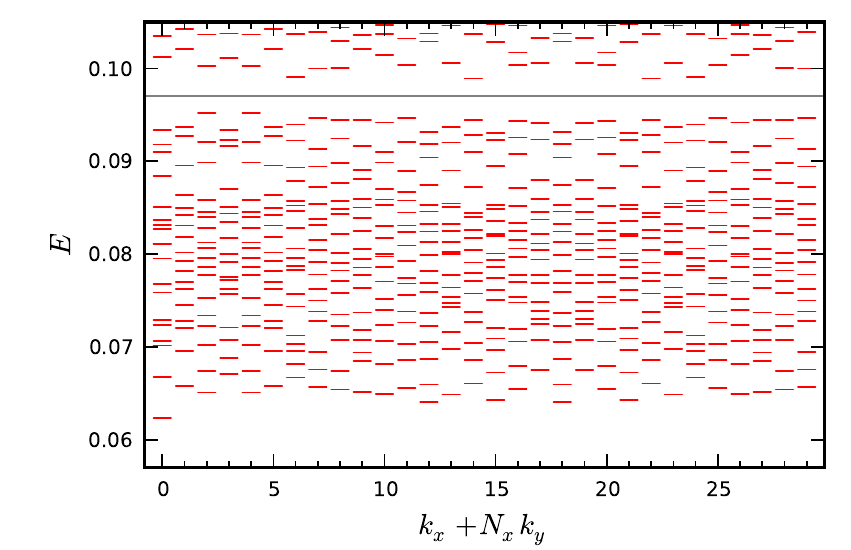}
\caption{\label{fig:965} Low energy spectrum of the Haldane model with $N=9$ 
particles on the
$N_x\times N_y=6\times 5$ lattice. The number of states below the gray line is 
$19$ in momentum sectors with $k_x=0,3$ and $18$ elsewhere, in agreement with 
the $(1,3)$-admissible counting rule.}
\end{figure}

Three-fold degeneracy alone is insufficient to fully establish the observed 
ground state as a FQH state; it could very well be an imprint of a lattice 
charge density wave.
To rule out this alternative we need to study the excitations of the 
system. We decrease the filling $\nu$ from $1/3$ and check for
quasiholes in the energy spectrum. In Fig.~\ref{fig:764} we show the 
energy spectrum of $N=7$ particles on the $N_x\times N_y=6\times 4$ lattice, 
and in Fig.~\ref{fig:965} the energy spectrum of $N=9$ particles on the
$N_x\times N_y=6\times 5$ lattice. 
(We run into convergence problems at $N=11$.)
These configurations have the same lattice 
geometry as the corresponding ground states shown earlier, but with one electron 
removed. An energy gap is clearly visible in the spectrum, and the low-energy 
part has the same counting in each momentum sector as predicted by 
the $(1,3)$-admissible rule. This further substantiates that the ground state 
observed at filling $1/3$ indeed has the basic features of the Laughlin FQH  
state~\cite{Laughlin83:Nobel}.

\subsection{Entanglement spectrum}
Recent developments~\cite{Li08:ES,Sterdyniak11:PES,Chandran11:ES} showed that the 
excitations in FQH systems are manifested in the entanglement between 
particles~\cite{Li08:ES,Zozulya07,Haque07:PEE,Haque09} in the many-body ground 
state. Using this alternative probe, we 
provide further evidence that the ground state at filling $1/3$ is a FQH 
Laughlin state. This tool is highly valuable in the present case since no 
overlap with model wave functions can be computed: despite several 
proposals~\cite{Qi11:Wavefunction,Lu11:Parton,McGreevy11:Parton,Vaezi11:Parton}, 
concrete expressions for the model wave functions have not been established 
for FCI.

\begin{figure}[]
\includegraphics{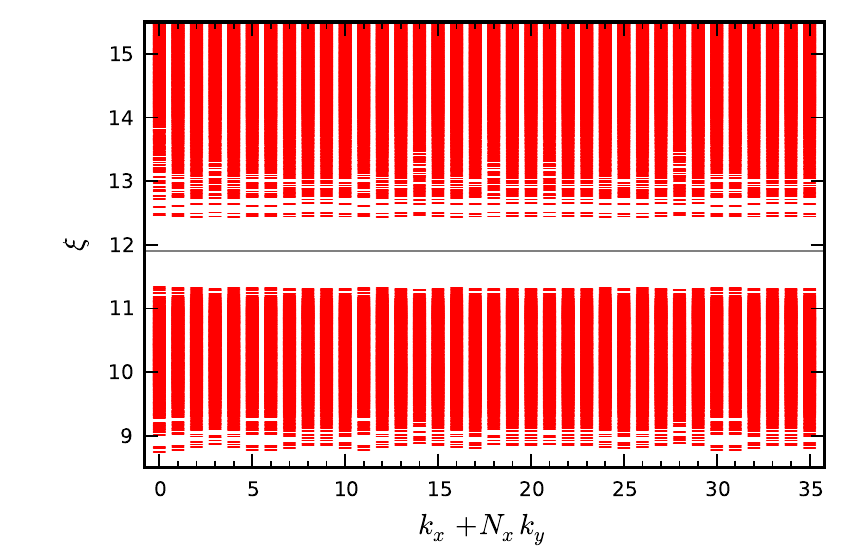}
\caption{\label{fig:1266pes} Particle entanglement spectrum of the ground 
state of the Haldane model of $N=12$ particles on the $N_x\times N_y=6\times 6$ 
lattice, with $N_B=8$ particles traced out. The number of states below the 
gray line is $741$ in momentum sectors where both $k_x$ and $k_y$ are even
and $728$ elsewhere, in agreement with the $(1,3)$-admissible counting rule.
The system size is exactly the same as Fig. 13 of 
Ref.~\onlinecite{Regnault11:FCI}.}
\end{figure}

Specifically, we cut the system in the way described in 
Ref.~\onlinecite{Sterdyniak11:PES} and further used to look at the FCIs 
in Ref.~\onlinecite{Regnault11:FCI}.
We divide $N$ particles into two subsystem of $N_A$ and $N_B$ 
particles, and trace out the degrees of freedom carried by the $N_B$ 
particles. The eigenvalues $e^{-\xi}$ of the resulting reduced density matrix 
$\rho_A=\text{Tr}_B\rho=\text{Tr}_B|\psi\rangle\langle\psi|$ define the
particle entanglement energies $\xi$. For degenerate ground states, we form 
the density matrix as an incoherent sum with equal weights~\cite{Regnault11:FCI} 
$\rho=\frac{1}{3}\sum_i |\psi_i\rangle\langle\psi_i|$. Then, the entanglement 
energy levels $\xi$ can be displayed in groups marked by the total momentum 
$(k_x,k_y)$ of the $N_A$ particles. 
A typical case is shown in Fig.~\ref{fig:1266pes}. The spectrum is very 
similar to what was found in 
the checkerboard lattice model~\cite{Regnault11:FCI}. We observe a clear, 
although narrower, entanglement gap in the spectrum; the counting of the 
entanglement energy levels below the gap matches (in each momentum sector) 
the $(1,3)$-admissible quasihole counting~\cite{Bernevig11:Counting} of $N_A$ 
particles on the $N_x\times N_y$ reciprocal lattice. 
We have checked all values of $N_A$ manageable by current computers and have 
found perfect agreement with the counting principle in all cases.
Given the vast difference between the checkerboard lattice model and the 
Haldane model (lattice symmetry, coordination number, flux distribution, 
etc.), the similarity in the entanglement spectrum is surprising. Compared 
with the relatively fuzzy quasihole energy spectrum, the ground state 
entanglement spectrum turns out to be a more reliable alternative route to 
probe the physics of fractional excitations. 

\subsection{Parameter dependence}\label{sec:haldane-paramdep}

As analyzed by Haldane~\cite{Haldane88:Honeycomb}, strong inversion 
breaking eventually overcomes the non-trivial topology; the Chern number vanishes
when $|M|>3\sqrt{3}|t_2\sin\phi|$. We now study whether strong interactions 
would further shrink the volume of the topologically non-trivial phase in 
parameter space. The density-density interaction projected to the flattened 
valence band has two parametric degrees of freedom $(M/t_1,t_2\sin\phi/t_1)$, 
which effectively change the Berry curvature in the model. 
Without loss of generality, we fix $t_1=t_2=1$ and study the region with $M>0$ 
and $\phi\in[0,\pi/2]$. There is no clear boundary between different 
interacting phases in finite-size numerics. We thus need a quantitative 
characterization of the similarity to the ideal FCI state. 

We denote the $(1,3)$-admissible counting of $N$ 
particles on the $N_x\times N_y$ lattice in momentum sector $(k_x,k_y)$ by 
$n^{N,N_x,N_y}_{k_x,k_y}$. We denote by $\mathcal{A}$ 
the collection of the lowest $n^{N,N_x,N_y}_{k_x,k_y}$ states in each momentum 
sector, and by $\bar{\mathcal{A}}$ the collection of all the other states. If 
the system is in a well-developed FCI state, the collection 
$\mathcal{A}$ is the 3-fold degenerate ground state, while the collection 
$\bar{\mathcal{A}}$ contains the excited states. The energy 
gap is thus 
$\Delta E=\min E_{\bar{\mathcal{A}}} - \max E_{\mathcal{A}}$,
and the energy spread of the ground-state manifold can be defined as 
$\delta E=\max E_{\mathcal{A}}-\min E_{\mathcal{A}}$. Further, we calculate the 
entanglement spectra of the degenerate ground state for various partitions 
$(N_A,N_B)$ of $N$ particles. For each $N_A$, we denote by $\mathbb{A}$ the 
collection of the lowest $n^{N_A,N_x,N_y}_{k_x,k_y}$ entanglement energy 
levels in each momentum sector, and by $\bar{\mathbb{A}}$ all the other 
entanglement energy levels. We define the entanglement gap 
$\Delta \xi=\min \xi_{\bar{\mathbb{A}}}-\max \xi_{\mathbb{A}}$.
The parameter sets with large $\Delta E$, small $\delta E$ and large 
$\Delta \xi$ are likely to host a FCI state.

\begin{figure}[]
\includegraphics{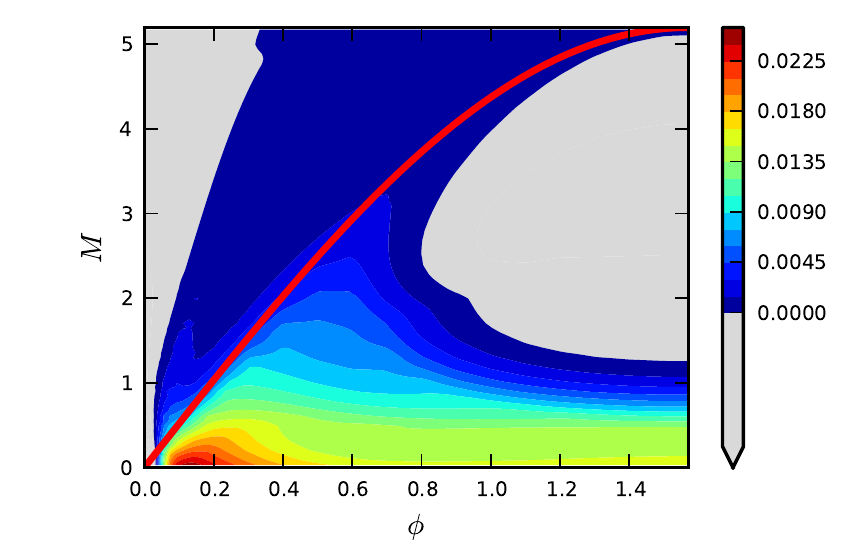}
\caption{\label{fig:haldane-gap-2d} The distribution of the energy gap $\Delta E$ 
in the $(\phi,M)$ plane for the Haldane model of $N=8$ particles on the
$N_x\times N_y=6\times 4$ lattice. The color code for $\Delta E$ is shown by 
the bar on the right, and the region with the gap closed is marked in gray. 
The bold red line $M=3\sqrt{3}\sin\phi$ separates the topologically 
non-trivial sector of the single-particle problem from the trivial one.}
\end{figure}

\begin{figure}[]
\includegraphics{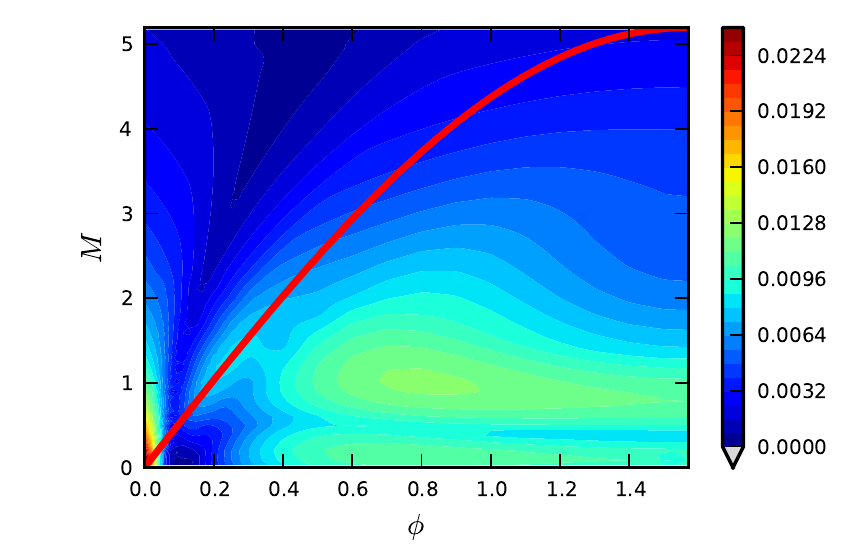}
\caption{\label{fig:haldane-spread-2d} The distribution of the energy spread 
$\delta E$ in the $(\phi,M)$ plane for the Haldane model of $N=8$ particles on 
the $N_x\times N_y=6\times 4$ lattice. The color code for $\delta E$ is 
shown by the bar on the right. 
The bold red line $M=3\sqrt{3}\sin\phi$ separates the topologically 
non-trivial sector of the single-particle problem from the trivial one.
}
\end{figure}

\begin{figure}[]
\includegraphics{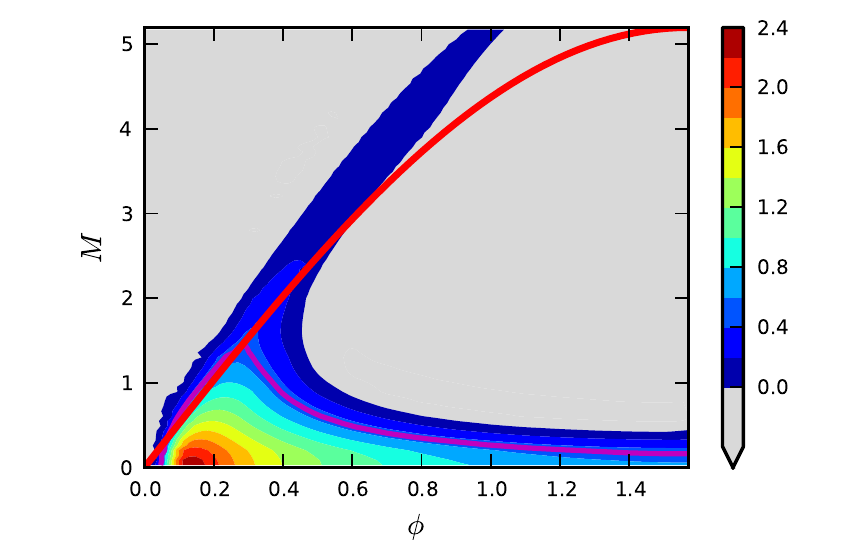}
\caption{\label{fig:haldane-entgap-2d} The distribution of the entanglement gap 
$\Delta \xi$ in the $(\phi,M)$ plane for the ground state of the Haldane model of 
$N=8$ particles on the $N_x\times N_y=6\times 4$ lattice, with $N_B=5$ particles 
traced out. 
The color code for $\Delta \xi$ is shown by 
the bar on the right, and the region with the gap closed is marked in gray.
The average spacing of the entanglement levels below the gap is 
at most $0.05$ (not shown in the figure); hence the entanglement gap should be 
considered widely open as long as $\Delta \xi\gg 0.05$. The line with 
$\Delta \xi=0.5$ is shown in magenta.
The bold red line $M=3\sqrt{3}\sin\phi$ separates the topologically 
non-trivial sector of the single-particle problem from the trivial one.
}
\end{figure}

In Figs.~\ref{fig:haldane-gap-2d}, \ref{fig:haldane-spread-2d}, 
\ref{fig:haldane-entgap-2d}, we show the distribution of $\Delta E$, $\delta E$, 
$\Delta \xi$ over the $(\phi,M)$ plane. Combining the three plots, we find 
that the region enclosed by $|M|<3\sqrt{3}|t_2\sin\phi|$ and 
$|M|\lesssim 3\sqrt{3}|t_2|(0.4-|\sin\phi|)$ has large $\Delta E$ and 
$\Delta \xi$, and small $\delta E$, and hence is likely to support a FCI phase.

\subsection{Berry curvature variation}

A few authors pointed out that a Chern band shares an important feature with 
the Landau level~\cite{Parameswaran11:W-inf,Murthy11:CF,Bernevig11:Counting}. 
In the limit of long wave-length and uniform Berry curvature, the projected 
density operators form a closed Lie algebra. This algebra has the same 
structure as the Girvin-MacDonald-Platzman algebra of magnetic 
translations~\cite{Girvin86:GMP}, with the Berry curvature taking the 
role of the uniform magnetic field. They thus argued that the development of the 
FCI phase is driven by this algebraic structure. This picture 
suggests a negative correlation between Berry curvature fluctuations and the 
propensity towards a FCI phase. This is indeed observed in finite-size 
numerics, as we describe below. 

\begin{figure}[]
\includegraphics{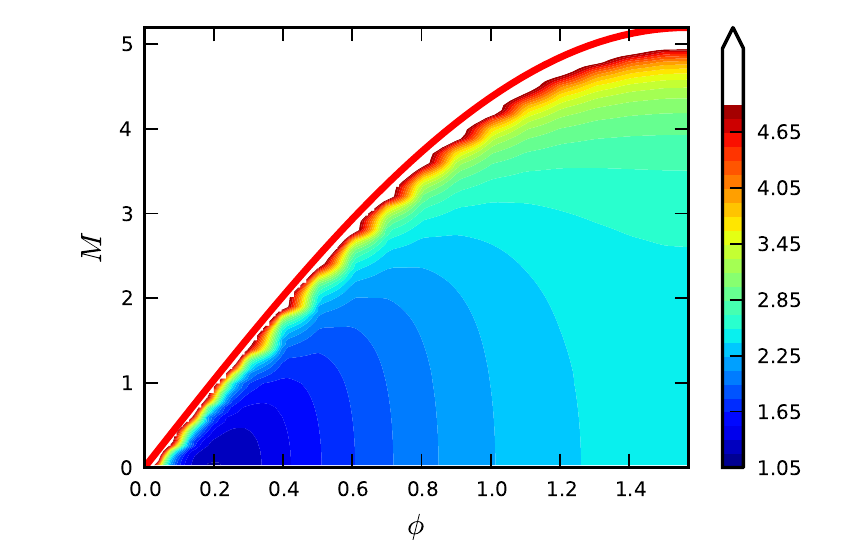}
\caption{\label{fig:haldane-bsd-2d} The distribution of the standard deviation 
of the Berry curvature $\sigma_{\mathcal{B}}$ in the $(\phi,M)$ plane for the 
Haldane model of $N=8$ particles on the $N_x\times N_y=6\times 4$ lattice. The 
bar on the right shows the color code for the value of 
$2\pi\sigma_{\mathcal{B}}$.
The bold red line $M=3\sqrt{3}\sin\phi$ separates the topologically 
non-trivial sector of the single-particle problem from the trivial one.
We focus on the non-trivial sector. The strong fluctuation of the Berry 
curvature very close to the boundary of the non-trivial sector is not shown. 
}
\end{figure}

We measure Berry curvature fluctuations by the simplest possible option, its 
standard deviation $\sigma_\mathcal{B}$ in units of the average Berry 
curvature $|\bar{\mathcal{B}}|=1/2\pi$. The distribution of 
$2\pi\sigma_{\mathcal{B}}$ is shown in Fig.~\ref{fig:haldane-bsd-2d}.
Comparing this with the patterns in Figs.~\ref{fig:haldane-gap-2d}, 
\ref{fig:haldane-spread-2d}, \ref{fig:haldane-entgap-2d}, we find a clear
correlation between $\sigma_{\mathcal{B}}$ and the three measures of the 
propensity towards the FCI phase over the full range of the parameter scan. In 
particular, the optimal parameter region supporting a robust FCI phase coincides 
with the part of the parameter space with least fluctuating Berry curvature. 
While this agrees with the picture we expect from the algebraic 
structure~\cite{Parameswaran11:W-inf,Murthy11:CF,Bernevig11:Counting,Goerbig11},
we stress that the fluctuation at the optimal parameters is still quite 
significant, with its standard deviation comparable to the mean value.
The FCI is apparently more robust to Berry curvature fluctuations than 
expected. This is crucial for two-band Chern insulators since their Berry 
curvature cannot be completely flat due to the no-hair 
theorem~\cite{Podolsky11:Unpublished}.

\section{Two-orbital Model}

\begin{figure}[]
\centering
\includegraphics[]{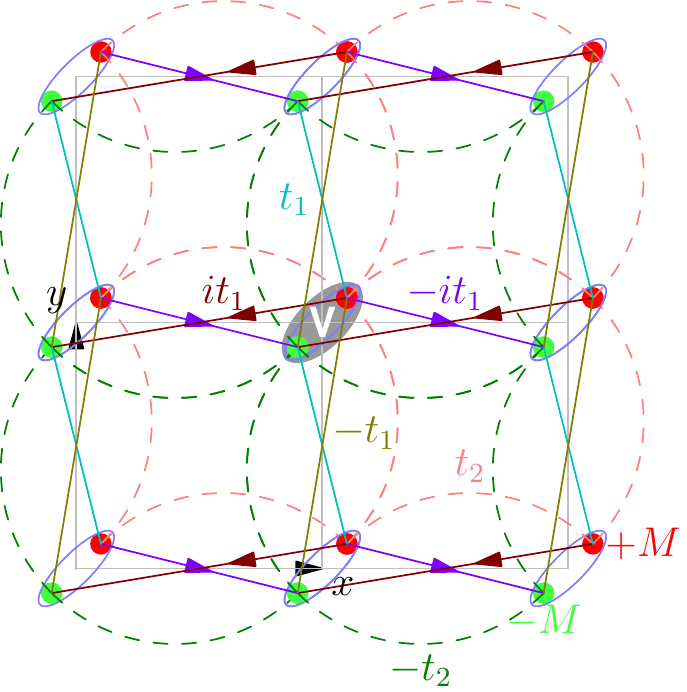}%
\caption{\label{fig:square} The two-orbital model on square lattice. The two 
orbitals $A$,$B$ are colored in red and green, respectively. The intra-orbital NN
hopping amplitude is $t_2$ between the $A$ orbitals, and $-t_2$ between the $B$ 
orbitals. The inter-orbital NN hopping amplitude is $\pm it_1$ in the $x$ 
direction along the arrows, and $\pm t_1$ in the $y$ direction.
The on-site Hubbard repulsion is depicted by the ellipse in gray.
}
\end{figure}

We now turn to a simpler model with two orbitals on each site of a square 
lattice. This model represents the spin-up half of the Mercury-Telluride 
two-dimensional topological insulator~\cite{Bernevig06:BHZ}. As shown in 
Fig.~\ref{fig:square}, there are two orbitals $A$, $B$ on each site, with 
energy difference $2M$. The intra-orbital NN hopping has amplitude $\pm t_2$, 
while the inter-orbital NN hopping has amplitude $\pm it_1$ and $\pm t_1$ in the 
$x$ and $y$ directions, respectively.
These amplitudes have been specifically designed such that the single-particle 
Bloch Hamiltonian takes the form 
$H=\sum_\mathbf{k} (\psi^\dagger_{\mathbf{k},A}, \psi^\dagger_{\mathbf{k},B})
h(\mathbf{k})(\psi_{\mathbf{k},A}, \psi_{\mathbf{k},B})^\text{T}$,
where
\begin{align}\nonumber
h(\mathbf{k})=\ & 2t_1\sin k_x\sigma_x + 2t_1\sin k_y\sigma_y \\
&+[M-2t_2(\cos k_x+\cos k_y)]\sigma_z.
\end{align}
Similar to the case of the Haldane model, we can flatten the Bloch bands using 
projectors. We then add Hubbard inter-orbital repulsion to each site, as shown 
in Fig.~\ref{fig:square}. In momentum space, the interaction term reads
\begin{equation}\label{eq:square-interaction}
\frac{1}{N}\sum_{\{\mathbf{k}_i\}} 
\delta_{\mathbf{k}_1+\mathbf{k}_2-\mathbf{k}_3-\mathbf{k}_4}^{\text{mod} 2\pi}
\psi^\dagger_{\mathbf{k}_3A}\psi^\dagger_{\mathbf{k}_4B}
\psi_{\mathbf{k}_2B}\psi_{\mathbf{k}_1A}.
\end{equation}
Even though at single-particle level a two-orbital-per-site model is 
equivalent to a model with two sites in each unit cell, this situation changes 
when interactions are included. Notice that the interaction here
(Eq.~\ref{eq:square-interaction}) has a different form factor than the one in 
Eq.~\ref{eq:haldane-interaction-factor}.

\begin{figure}[]
\includegraphics{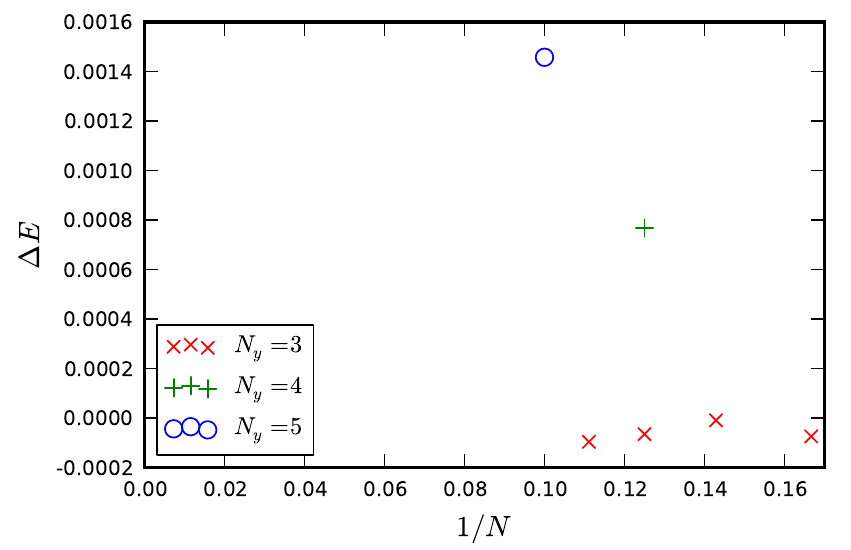}
\caption{\label{fig:square-scaling} Energy gap $\Delta E$ of the two-orbital 
model for different system sizes and aspect ratios. In each case, $N_x=3N/N_y$.
The negative values of $\Delta E$ in the $N_y=3$ group mean that the actual 
ground state is not in the momentum sectors predicted by the theoretical 
counting rule.
}
\end{figure}

\begin{figure}[]
\includegraphics{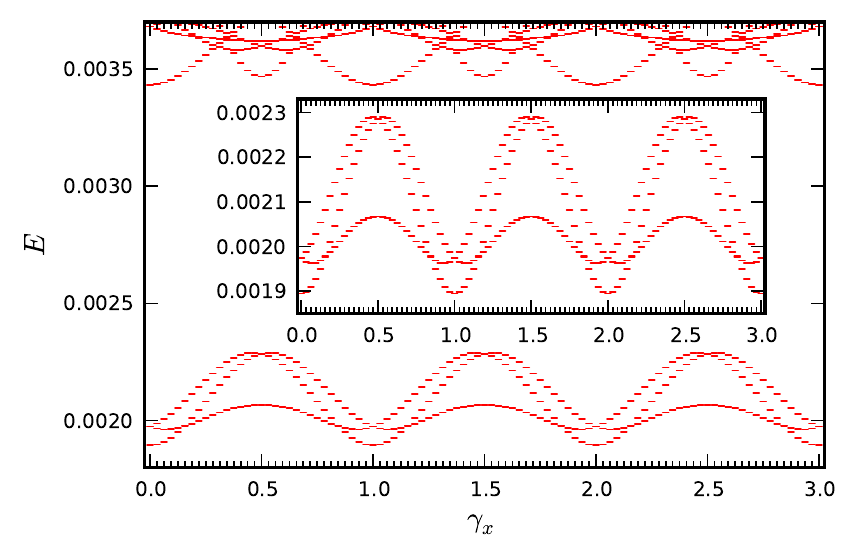}
\caption{\label{fig:square-1065twisted} Spectral flow of the low-lying
states of the two-orbital model with $N=10$ particles on the
$N_x\times N_y=6\times 5$ lattice upon flux insertion along the $x$ direction. 
$\gamma_x$ counts the number of fluxes inserted.
The 3-fold ground states flow into each other, and do not mix with higher 
states during flux insertion. It takes 3 full fluxes for the 3-fold states to 
return to the original configuration (inset).}
\end{figure}

\begin{figure}[]
\includegraphics{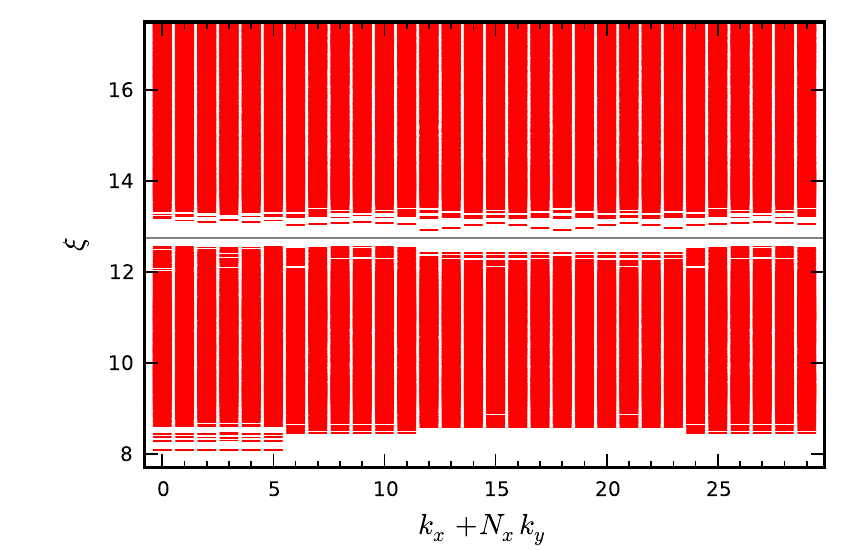}
\caption{\label{fig:square-1065pes} Particle entanglement spectrum of the ground 
state of the two-orbital model of $N=10$ particles on the 
$N_x\times N_y=6\times 5$ lattice, with $N_B=5$ particles traced out.
The number of states below the gray line is $776$ in momentum sectors with 
$k_y=0$ and $775$ in all the other sectors, in agreement with the analytical 
counting rule.}
\end{figure}

We diagonalize the interacting Hamiltonian in the flattened lowest band at 
filling $1/3$ and $(t_2/t_1,M/t_1)=(1,2)$.
We show the energy spectrum of $N=8,10$ particles on the
$N_x\times N_y=6\times \frac{N}{2}$ lattice in 
Fig.~\ref{fig:square-864,1065}. 
(We run into serious convergence problems at $N=12$.)
In the two cases ($N=8,10$), a 3-fold degenerate 
ground state is seen at total momenta $\{(0,0),(2,0),(4,0)\}$ and
$\{(1,0),(3,0),(5,0)\}$, respectively. This 
agrees perfectly with the $(1,3)$-admissible counting proposed in 
Ref.~\onlinecite{Regnault11:FCI,Bernevig11:Counting}. 
As shown in Fig.~\ref{fig:square-scaling},
the energy gap $\Delta E$ remains open and scales to a finite value in the 
limit of $N\rightarrow\infty$ with $N_x/N_y$ finite. 
The three degenerate ground states exhibit 
spectral flow upon flux insertion with a period of 3 fluxes (see 
Fig.~\ref{fig:square-1065twisted}). This shows that the system has Hall 
conductance $\sigma_{xy}=1/3$. 
We study the quasihole excitations through ground state entanglement. In 
Fig.~\ref{fig:square-1065pes}, we observe an entanglement gap in the spectrum. 
The counting of the entanglement energy levels below the gap again matches 
in each momentum sector the $(1,3)$-admissible counting. This shows that the 
excitations in the ground state of the two-orbital model have the same 
counting as that of Abelian fractional statistics $1/3$ quasiholes. We thus 
conclude that the ground state is a FQH Laughlin state. 

\begin{figure}[]
\includegraphics{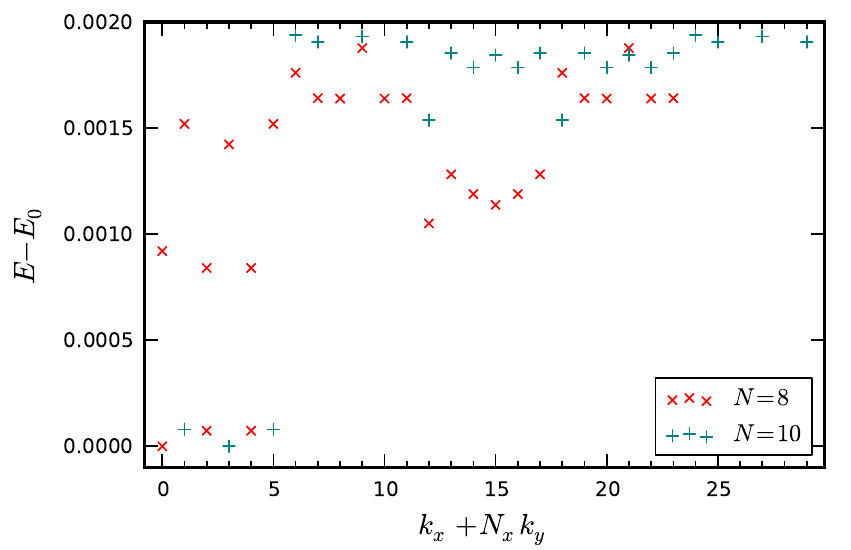}
\caption{\label{fig:square-864,1065} Low energy spectrum of two-orbital model
for $N=8$ (marked by red crosses) and $N=10$ (green plus signs) 
particles on $(N_x,N_y)=(6,N/2)$ lattice, with energies shifted by $E_0$, the 
lowest energy for each system size. We only show the lowest excited level in 
each momentum sector in addition to the 3-fold ground state.
}
\end{figure}

\begin{figure}[]
\includegraphics{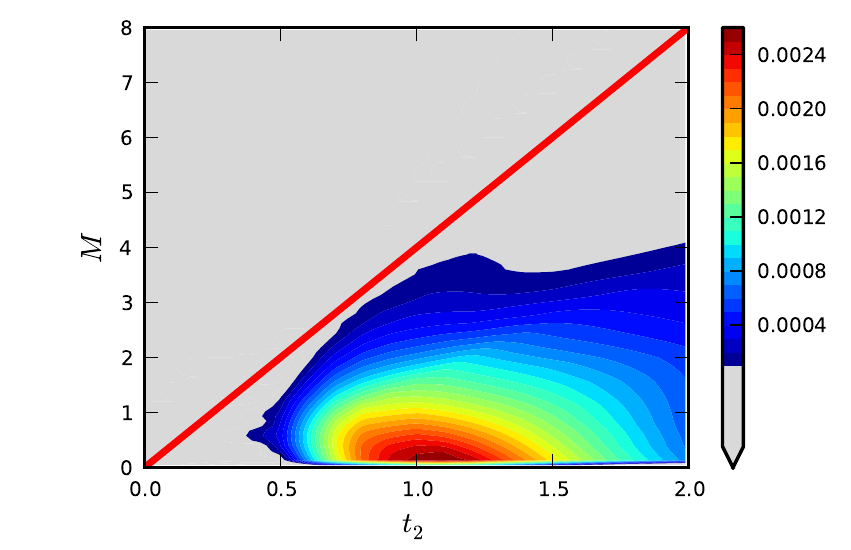}
\caption{\label{fig:square-gap-2d} The distribution of the energy gap $\Delta E$ 
in the $(t_2,M)$ plane for the two-orbital model of $N=8$ particles on the
$N_x\times N_y=6\times 4$ lattice. The color code for $\Delta E$ is shown by 
the bar on the right, and the region with the gap closed is marked in gray. 
The bold red line $M=4t_2$ separates the topologically 
non-trivial sector of the single-particle problem from the trivial one.
}
\end{figure}

\begin{figure}[]
\includegraphics{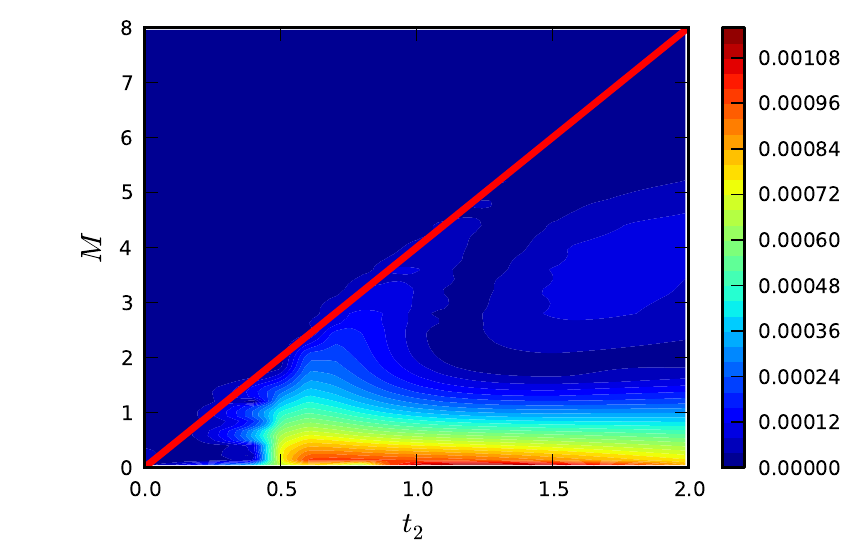}
\caption{\label{fig:square-spread-2d} The distribution of the energy spread $\delta E$ 
in the $(t_2,M)$ plane for the two-orbital model of $N=8$ particles on the
$N_x\times N_y=6\times 4$ lattice. 
The color code for $\delta E$ is shown by the bar on the right. 
The bold red line $M=4t_2$ separates the topologically 
non-trivial sector of the single-particle problem from the trivial one.
}
\end{figure}

\begin{figure}[]
\includegraphics{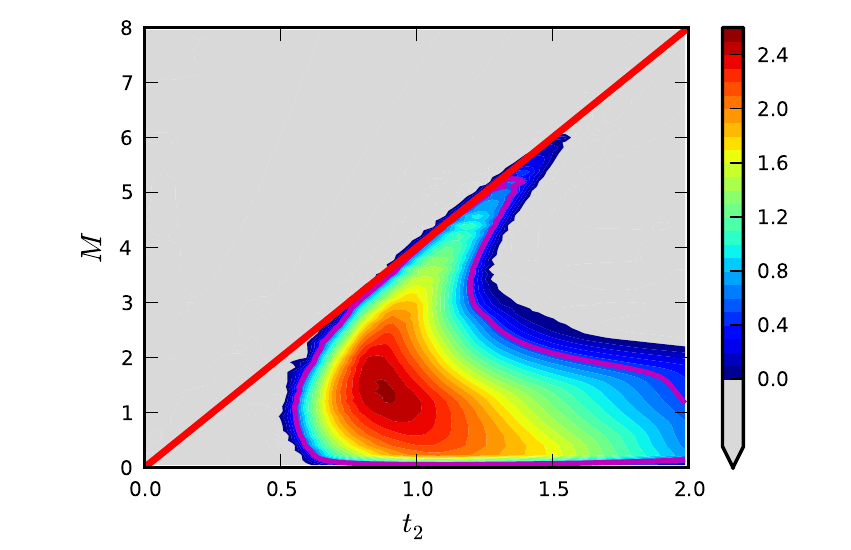}
\caption{\label{fig:square-entgap-2d} The distribution of the entanglement gap 
$\Delta \xi$ in the $(t_2,M)$ plane for the ground state of the two-orbital model 
of $N=8$ particles on the $N_x\times N_y=6\times 4$ lattice, with $N_B=5$ 
particles traced out. 
The color code for $\Delta \xi$ is shown by 
the bar on the right, and the region with the gap closed is marked in gray.
The average spacing of the entanglement levels below the gap is 
at most $0.05$ (not shown in the figure); hence the entanglement gap should be 
considered widely open as long as $\Delta \xi\gg 0.05$. The line with
$\Delta \xi=0.5$ is shown in magenta. 
The bold red line $M=4t_2$ separates the topologically 
non-trivial sector of the single-particle problem from the trivial one.
}
\end{figure}

\begin{figure}[]
\includegraphics{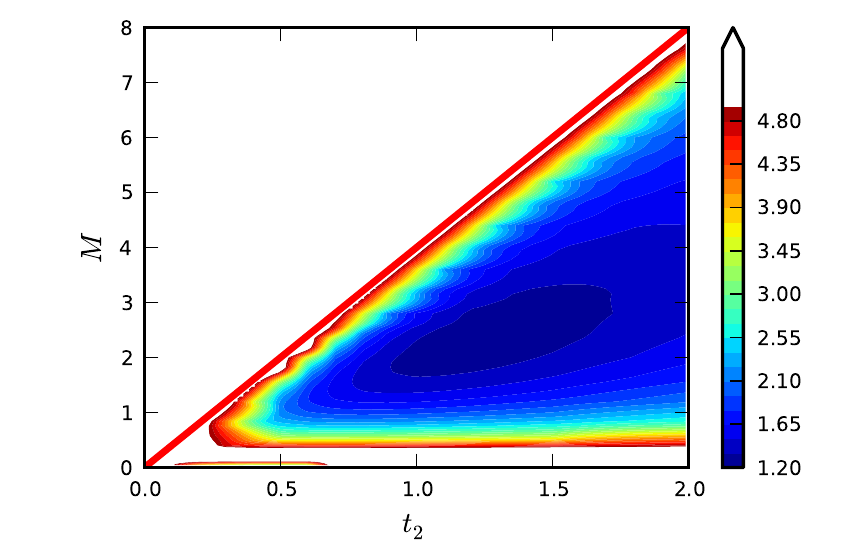}
\caption{\label{fig:square-bsd-2d} The distribution of the standard deviation 
of the Berry curvature $\sigma_{\mathcal{B}}$ in the $(t_2,M)$ plane for the 
two-orbital model of $N=8$ particles on the $N_x\times N_y=6\times 4$ lattice. 
The bar on the right shows the color code for the value of 
$2\pi\sigma_{\mathcal{B}}$. 
The bold red line $M=4t_2$ separates the topologically 
non-trivial sector of the single-particle problem from the trivial one.
We focus on the non-trivial sector. The strong fluctuation of the Berry 
curvature very close to the boundary of the non-trivial sector is not shown. 
}
\end{figure}

The two-orbital model has topologically non-trivial bands when $|M|<4|t_2|$. 
We now move the system away from the point $(t_2,M)=(1,2)$ and probe this 
parameter region. In Figs.~\ref{fig:square-gap-2d}, 
\ref{fig:square-spread-2d}, \ref{fig:square-entgap-2d}, we display the 
distribution of the energy gap $\Delta E$, the energy spread $\delta E$, and 
the entanglement gap $\Delta \xi$ of the ground state. Compared with the 
Haldane model, the situation is more complicated. The maximum of the energy 
gap $\Delta E$ does not coincide with the minimum of the energy spread 
$\delta E$. Rather, the region with large gap $\Delta E$ 
tends to have large spread $\delta E$ as well. The peak of the entanglement gap 
$\Delta \xi$ is close neither to the maximum of $\Delta E$ nor to the minimum 
of $\delta E$. Varying $(M,t_2)$ scans over the manifold of the two-orbital 
interacting Hamiltonians projected to the lowest band. The lack of correlation 
between $\Delta E$, $\delta E$ and $\Delta \xi$ 
suggests that the distance from this manifold of Hamiltonians to the Laughlin 
model Hamiltonian is quite large. 
Further, we check the correlation between Berry curvature fluctuations and the 
propensity towards the FCI phase. The variation of the curvature fluctuation 
is shown in Fig.~\ref{fig:square-bsd-2d}. We observe weak correlation 
between $\sigma_\mathcal{B}$ and $\delta E$ and find little association between 
$\sigma_\mathcal{B}$ and $\Delta E$ or $\Delta \xi$. It is possible that 
$\sigma_\mathcal{B}$ is not a very good measure of the Berry curvature 
fluctuation. We note that in this model the interaction before projection is 
constant. We leave in-depth study of this issue for future work.

\section{Kagome Lattice Model}

\begin{figure}[]
\centering
\includegraphics{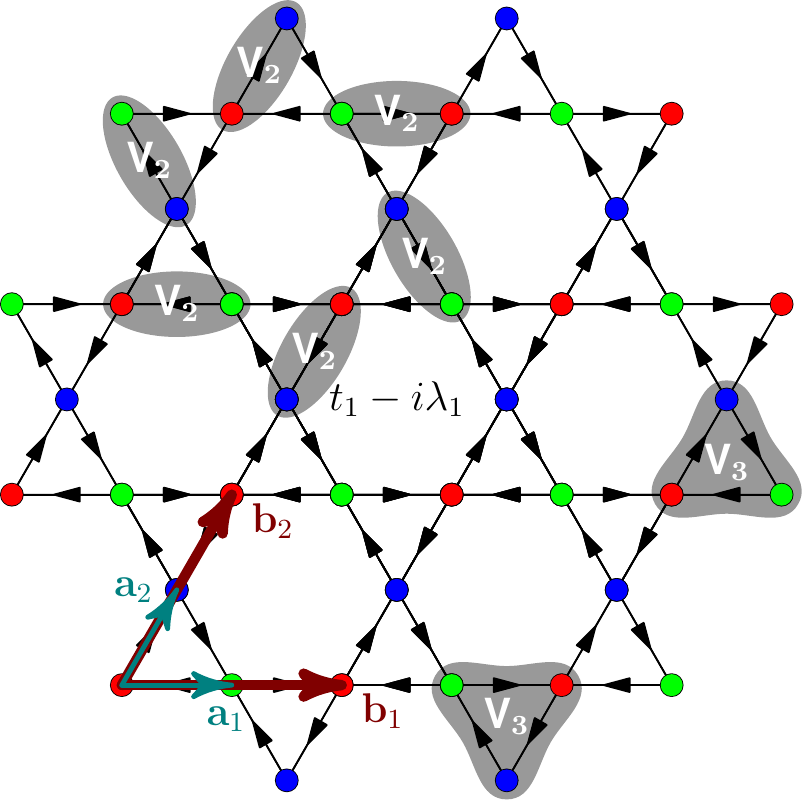}%lbrt
\caption{\label{fig:kagome} The Kagome lattice model with NN spin-orbit 
coupling. The three sublattices $A$,$B$,$C$ are colored in red, green, blue, 
respectively. The lattice translation vectors are $\mathbf{b}_1=2\mathbf{a}_1$, 
$\mathbf{b}_2=2\mathbf{a}_2$. The hopping amplitude between NN is 
$t_1-i\lambda_1$ in the direction of arrows.
The two-body and the three-body interactions between NN are illustrated by the 
gray ellipses and triangles, respectively.
}
\end{figure}

We study the Kagome lattice model built by Tang et al.~\cite{Tang11:Kagome}. 
As shown in Fig.~\ref{fig:kagome}, the lattice is spanned by the translation 
vectors $\mathbf{b}_1$ and $\mathbf{b}_2$, and it consists of three 
sublattices $A$,$B$,$C$. In Ref.~\onlinecite{Tang11:Kagome}, the 
single-particle model has 
been studied both without and with NNN hopping terms. We have looked at the 
effect of interactions in both cases. But for sake of simplicity we focus 
mostly on the case without NNN hoppings. The hopping amplitude between NN is 
$t_1\pm i\lambda_1$.
After a Fourier transform and a gauge transform 
$\psi_{\mathbf{k},B}\rightarrow\psi_{\mathbf{k},B}
e^{-i\mathbf{k}\cdot\mathbf{b}_1/2}$,
$\psi_{\mathbf{k},C}\rightarrow\psi_{\mathbf{k},C}
e^{-i\mathbf{k}\cdot\mathbf{b}_2/2}$,
the single-particle Hamiltonian can be cast in Bloch form as 
$H=\sum_\mathbf{k} (\psi^\dagger_{\mathbf{k},A}, \psi^\dagger_{\mathbf{k},B},
\psi^\dagger_{\mathbf{k},C})
h(\mathbf{k})(\psi_{\mathbf{k},A}, \psi_{\mathbf{k},B},
\psi_{\mathbf{k},C})^\text{T}$. 
Here the lattice momentum 
$\mathbf{k}=(\mathbf{k}\cdot \mathbf{b}_1,\mathbf{k}\cdot \mathbf{b}_2)
\equiv (k_x,k_y)$ is summed over the first Brillouin zone, and the $h(\mathbf{k})$ 
matrix reads
\begin{align}\nonumber
h(\mathbf{k})&=-t_1\left[
\begin{array}{ccc}
0 & 1+e^{-ik_x } & 1+e^{-ik_y } \\
1+e^{ik_x } & 0 & 1+e^{i(k_x -k_y )} \\
1+e^{ik_y } & 1+e^{i(k_y -k_x )} & 0
\end{array}
\right]\\
&+i\lambda_1\left[
\begin{array}{ccc}
0 & 1+e^{-ik_x } & -1-e^{-ik_y } \\
-1-e^{ik_x } & 0 & 1+e^{i(k_x -k_y )} \\
1+e^{ik_y } & -1-e^{i(k_y -k_x )} & 0
\end{array}
\right].
\end{align}
The three Bloch bands can be flattened using the projector method detailed for 
the Haldane model in Section~\ref{sec:haldane}. We focus on the lowest band. 
Unless specified otherwise, the numerical calculations shown below are 
performed at $\lambda_1=t_1$ as discussed in the original 
paper~\cite{Tang11:Kagome}. The lowest band has unit Chern number. 

\subsection{Filling $1/3$}

We fill the flattened lowest band to filling $1/3$, and add density-density 
repulsion between NN. After the gauge transform, the interaction term reads
\begin{equation}
\frac{1}{N}\sum_{\{\mathbf{k}_i\}}
\delta_{\mathbf{k}_1+\mathbf{k}_2-\mathbf{k}_3-\mathbf{k}_4}^{\text{mod} 2\pi}
\sum_{\alpha<\beta}^{A,B,C}
\psi^\dagger_{\mathbf{k}_3\alpha}\psi^\dagger_{\mathbf{k}_4\beta}
\psi_{\mathbf{k}_2\beta}\psi_{\mathbf{k}_1\alpha}
V^{\alpha\beta}_{\mathbf{k}_1\mathbf{k}_2\mathbf{k}_3\mathbf{k}_4},
\end{equation}
where the sublattice indices $(\alpha,\beta)$ are summed over $(A,B)$, $(B,C)$, 
$(C,A)$, and the interaction factors are
\begin{align}
V^{AB}_{\mathbf{k}_1\mathbf{k}_2\mathbf{k}_3\mathbf{k}_4}&= 1+
e^{-i(\mathbf{k}_2-\mathbf{k}_4)\cdot\mathbf{b}_1},\\\nonumber
V^{BC}_{\mathbf{k}_1\mathbf{k}_2\mathbf{k}_3\mathbf{k}_4}&= 1+
e^{i(\mathbf{k}_2-\mathbf{k}_4)\cdot(\mathbf{b}_1-\mathbf{b}_2)},\\\nonumber
V^{CA}_{\mathbf{k}_1\mathbf{k}_2\mathbf{k}_3\mathbf{k}_4}&= 1+
e^{i(\mathbf{k}_2-\mathbf{k}_4)\cdot\mathbf{b}_2}.
\end{align}
The 6 terms are illustrated by the 6 ellipses in Fig.~\ref{fig:kagome}.

\begin{figure}[]
\includegraphics{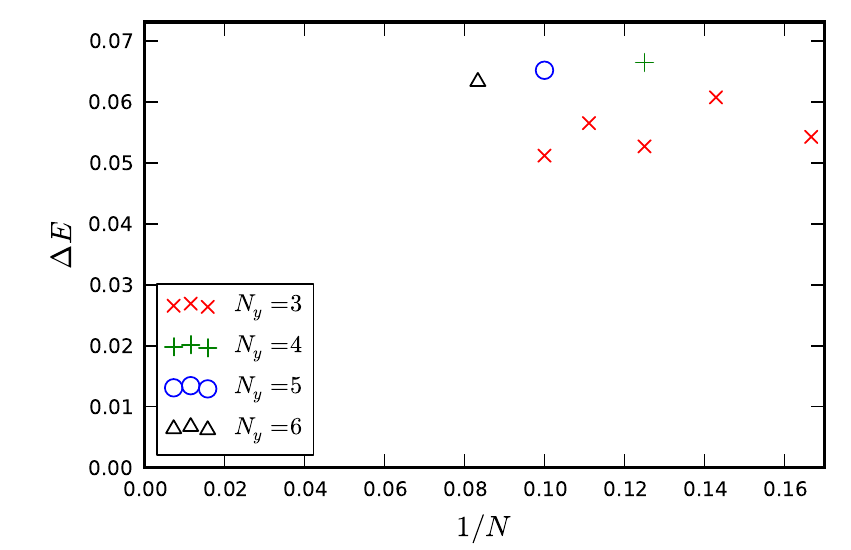}
\caption{\label{fig:kagome-scaling} Energy gap $\Delta E$ of the Kagome lattice 
model for different system sizes and aspect ratios. In each case, $N_x=3N/N_y$.}
\end{figure}

\begin{figure}[]
\includegraphics{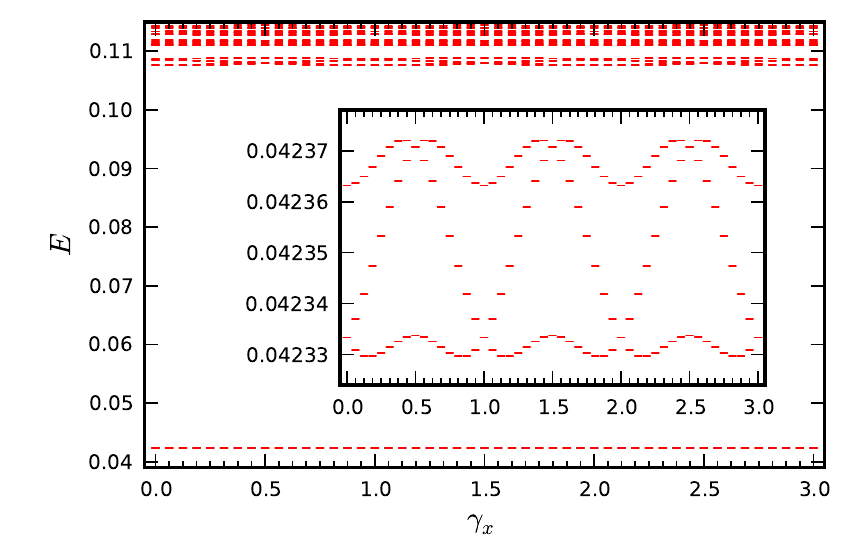}
\caption{\label{fig:kagome-1065twisted} Spectral flow of the low-lying
states of the Kagome lattice model with $N=10$ particles on the
$N_x\times N_y=6\times 5$ lattice upon flux insertion along the $x$ direction. 
$\gamma_x$ counts the number of fluxes inserted.
The 3-fold ground states flow into each other, and do not mix with higher 
states during flux insertion. It takes 3 full fluxes for the 3-fold states to 
return to the original configuration (inset).}
\end{figure}

We show the energy spectrum of $N=8,10,12$ particles on the
$N_x\times N_y=6\times \frac{N}{2}$ lattice in 
Fig.~\ref{fig:kagome-864,1065,1266}. 
In the three cases, a 3-fold degenerate 
ground state is seen at total momenta $\{(0,0),(2,0),(4,0)\}$, 
$\{(1,0),(3,0),(5,0)\}$, $\{(0,0),(0,0),(0,0)\}$, respectively. Again, this 
agrees perfectly with the $(1,3)$-admissible counting proposed in 
Ref.~\onlinecite{Regnault11:FCI,Bernevig11:Counting}. The ratio of the gap to 
the energy spread of the ground-state manifold is larger than that of the 
Haldane model. As shown in Fig.~\ref{fig:kagome-scaling},
the energy gap $\Delta E$ 
remains open and scales to a finite value in the 
limit of $N\rightarrow\infty$ with $N_x/N_y$ finite. 
And the three degenerate ground states exhibit spectral flow upon flux insertion. 
The period of 3 fluxes, shown in Fig.~\ref{fig:kagome-1065twisted}, 
indicates the system has Hall conductance $\sigma_{xy}=1/3$. 

\begin{figure}[]
\includegraphics{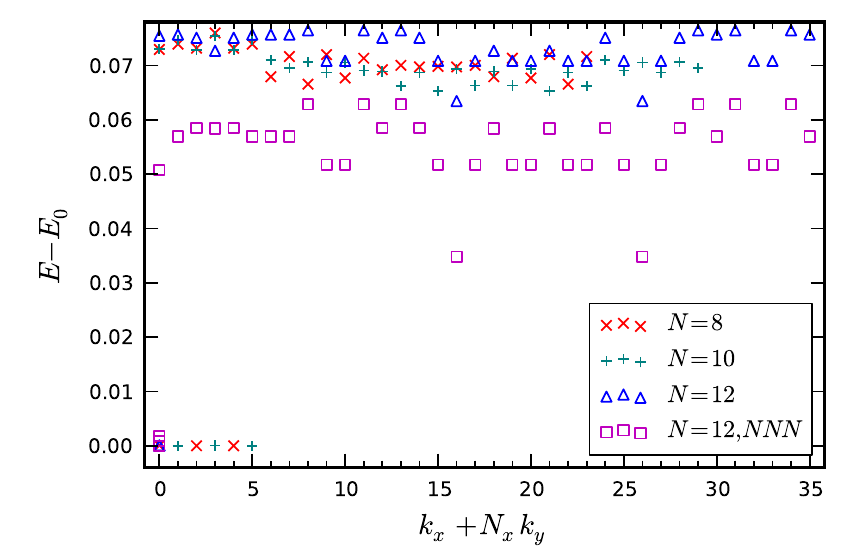}
\caption{\label{fig:kagome-864,1065,1266} Low energy spectrum of the Kagome 
lattice model with $N=8$ (marked by red crosses), $N=10$ (green plus signs), and $N=12$ 
(blue triangles) particles on the $N_x\times N_y=6\times N/2$ 
lattice, with energies shifted by $E_0$, the lowest energy for each system 
size. Also shown (magenta squares) is the spectrum of $N=12$ particles on 
the $N_x\times N_y=6\times 6$ lattice in the alternative Kagome model with NNN 
hoppings at the parameters suggested in the original 
paper~\cite{Tang11:Kagome}, namely $(\lambda_1,\lambda_2,t_2)=(0.28,0.2,-0.3)$.}
\end{figure}

\begin{figure}[]
\includegraphics{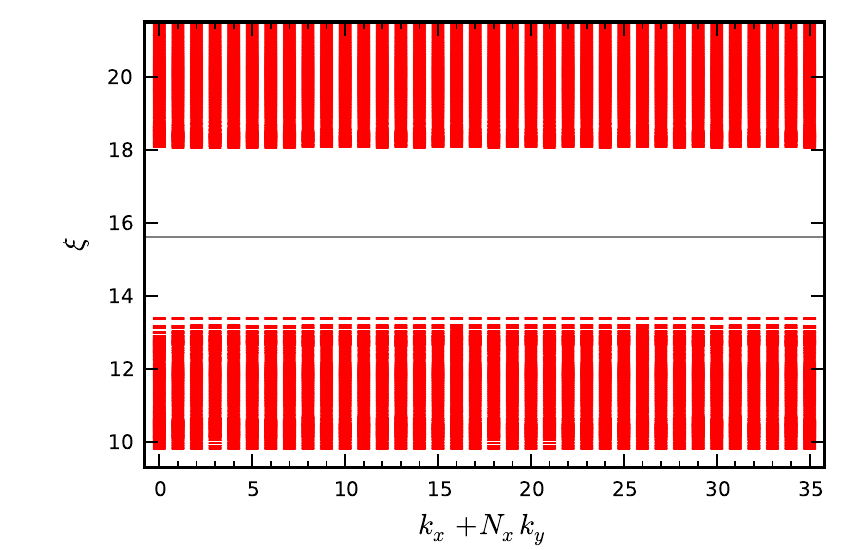}
\caption{\label{fig:kagome-1266pes} Particle entanglement spectrum of the ground 
state of the Kagome lattice model of $N=12$ particles on the 
$N_x\times N_y=6\times 6$ 
lattice, with $N_B=7$ particles traced out. The number of states below the 
gray line is $2530$ in each momentum sector, in agreement with the $(1,3)$-admissible 
counting rule. The width of the entanglement gap 
is $\Delta \xi=4.64$, meaning that spurious, non-FQH correlations have a 
probability of $e^{-4.64}\approx 0.0097$ relative to the universal ones.}
\end{figure}

We probe the quasihole excitations by the particle entanglement spectrum 
of the ground state. In Fig.~\ref{fig:kagome-1266pes}, we observe a clear, 
\emph{very large} entanglement gap in the spectrum, and the counting of 
the entanglement energy levels below the gap again matches in each momentum sector the 
$(1,3)$-admissible counting~\cite{Regnault11:FCI,Bernevig11:Counting}. The 
width of the entanglement gap is 
$\Delta \xi=4.64$. This means that in the ground state of the Kagome lattice 
model, the inter-particle correlations that obey the $(1,3)$ generalized Pauli 
principle is stronger than any other correlations \emph{by two orders of 
magnitude!} We thus conclude that the ground state is a FCI state with 
characteristics of the FQH Laughlin state.

\begin{figure}[]
\includegraphics{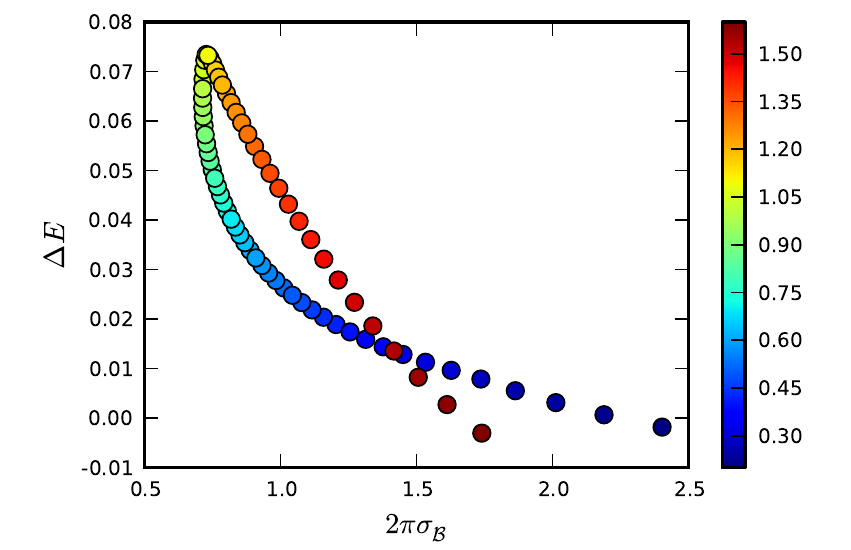}
\caption{\label{fig:kagome-gap-bsd} The correlation between the standard 
deviation $\sigma_\mathcal{B}$ of Berry curvature and the gap $\Delta E$ in 
the energy spectrum of 8 particles on the $N_x\times N_y=6\times 4$ lattice. The 
calculations are performed at $t_1=1$. The color of each scatter point encodes 
the value of $\lambda_1$ per the color bar to the right. A negative 
correlation between $\sigma_\mathcal{B}$ and $\Delta E$ is visible.}
\end{figure}

Now we briefly address the issue of parameter dependence. Without loss of 
generality, we fix $t_1=1$ and vary $\lambda_1$ in the range $(0,\sqrt{3})$. 
In this region, the single-particle spectrum is gapped and the lowest band 
has unit Chern number~\cite{Tang11:Kagome}. A strong correlation between the 
energy gap $\Delta E$ and the Berry curvature fluctuation 
$2\pi\sigma_\mathcal{B}$ on $\lambda_1$ is clearly visible in 
Fig.~\ref{fig:kagome-gap-bsd}. 

As a sanity check, we have also looked at the case where one partially fills 
the second band. We assume the lowest band is completely filled and inert, and 
diagonalize the second band directly. In such a situation, there is no 
evidence for Laughlin-like physics at filling $1+1/3$. This is expected since 
the second band has zero Chern number~\cite{Tang11:Kagome}.

\begin{figure}[]
\includegraphics{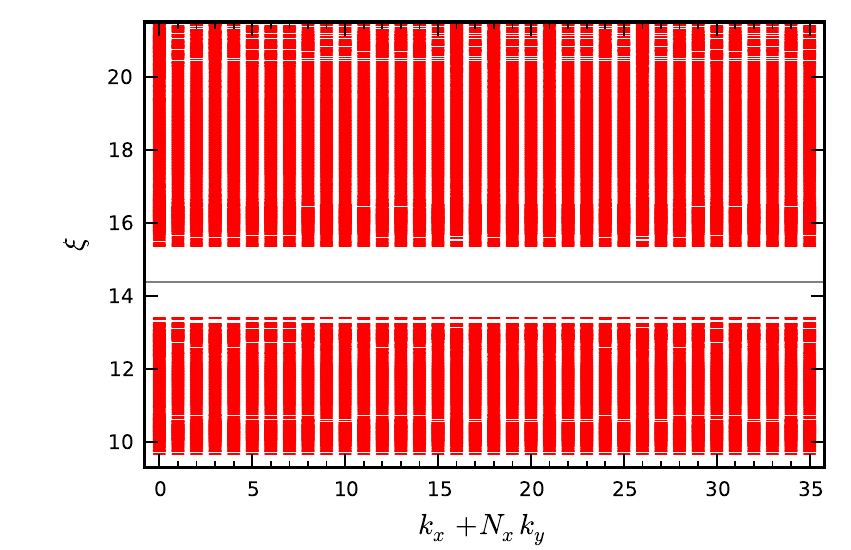}
\caption{\label{fig:kagome-1266pes-NNN} Particle entanglement spectrum of the 
ground state of the Kagome lattice model with NNN hoppings of $N=12$ particles 
on the $N_x\times N_y=6\times 6$ lattice at 
$(\lambda_1,\lambda_2,t_2)=(0.28,0.2,-0.3)$, with $N_B=7$ particles traced 
out. The number of states below the 
gray line is $2530$ in each momentum sector, in agreement with the $(1,3)$ 
counting rule. The width of the entanglement gap 
is $\Delta \xi=1.95$, much smaller than the gap of the model with only NN 
hoppings shown in Fig.~\ref{fig:kagome-1266pes}.}
\end{figure}

In Ref.~\onlinecite{Tang11:Kagome}, it was shown that by adding properly 
chosen NNN hopping terms, the band gap to bandwidth ratio can be enhanced by 
more than an order of magnitude, and it was argued that this alternative model 
could support a more robust Laughlin-like phase when interactions are added. 
Unfortunately, we observe the opposite effect: compared with the simple model 
with only NN hoppings, both the energy gap and the 
entanglement gap are only about half as large for the two cases 
with NNN hoppings studied in Ref.~\onlinecite{Tang11:Kagome}. 
The reduction of $\Delta\xi$ from $4.64$ to $1.95$ means the 
relative strength of the $(1,3)$ Pauli principle exclusion is reduced by a 
factor of $\sim 15$.
The energy and the entanglement spectra for the case with the largest band gap 
to bandwidth ratio [at $(\lambda_1,\lambda_2,t_2)=(0.28,0.2,-0.3)$] are shown in 
Figs.~\ref{fig:kagome-864,1065,1266} and~\ref{fig:kagome-1266pes-NNN}.
We have checked 
the nearby parameter region and no qualitative change is observed. Carefully tuning 
the NNN hoppings could achieve $\sim 10\%$ flatter Berry curvature, but the 
energy gap and the entanglement gap still end up smaller than the simple model 
with only NN hoppings, showing that the flat Berry curvature -- FQH phase 
correspondence is to be taken as a general trend, rather than a quantitative result. 

\subsection{Half filling}\label{sec:kagome-half-filling}

Among the FQH phases, the Moore-Read (MR) state plays a special role. It is one 
of the best candidates to explain the experimentally observed fraction $\nu=5/2$ 
and its excitations obey non-Abelian statistics. The $\mathbb{Z}_n$ 
Read-Rezayi (RR) states are the generalization of Laughlin $(n=1)$ and MR 
$(n=2)$ states. They occur at a filling factor $\nu=n/(n+2)$. The $\mathbb{Z}_3$ 
RR state is a potential candidate to explain the fraction $\nu=12/5$.

It has been recently shown that a $(n+1)$-body short-range interaction can 
stabilize the analogue of a $\mathbb{Z}_n$ RR state on the checkerboard 
lattice~\cite{Bernevig11:Counting}. The $(n+1)$-body interaction used 
in Ref.~\onlinecite{Bernevig11:Counting} mimics the short range interaction for which 
the $\mathbb{Z}_n$ RR state is the exact zero-energy ground state with highest 
density. It has also been very recently demonstrated~\cite{Wang11:MR} that the 
Haldane model of hard-core bosons exhibits a MR state at filling $\nu=1$. The 
three-body hard-core repulsion in that case was implemented by restricting the 
occupancy of any site to be less than three.
 
In this section, we focus on the half-filling case and thus on the fermionic 
MR state. Until now, we have found no evidence that a two-body NN 
interaction could stabilize a MR-like state in any of the lattice 
models studied. Instead of the two-body NN repulsion, we add to the Kagome 
lattice model a \emph{three}-body NN repulsion similar to the one in 
Ref.~\onlinecite{Bernevig11:Counting}, as shown by the gray triangles 
in Fig.~\ref{fig:kagome}. After the gauge transform, the 
density-density-density interaction term reads
\begin{align}\nonumber
\frac{1}{N}\sum_{\{\mathbf{k}_i\}}
\delta_{\substack{\mathbf{k}_1+\mathbf{k}_2+\mathbf{k}_3\\
-\mathbf{k}_4-\mathbf{k}_4-\mathbf{k}_6}}^{\text{mod} 2\pi}&
\psi^\dagger_{\mathbf{k}_4A}\psi^\dagger_{\mathbf{k}_5B}\psi^\dagger_{\mathbf{k}_6C}
\psi_{\mathbf{k}_3C}\psi_{\mathbf{k}_2B}\psi_{\mathbf{k}_1A}\\
&\times V_{\mathbf{k}_1\mathbf{k}_2\mathbf{k}_3\mathbf{k}_4\mathbf{k}_5\mathbf{k}_6},
\end{align}
where 
\begin{equation}
V_{\mathbf{k}_1\mathbf{k}_2\mathbf{k}_3\mathbf{k}_4\mathbf{k}_5\mathbf{k}_6}=
1+e^{-i(\mathbf{k}_2-\mathbf{k}_5)\cdot\mathbf{b}_1-i(\mathbf{k}_3-\mathbf{k}_6)\cdot\mathbf{b}_2}.
\end{equation}

It is well known that a short-range three-body repulsion stabilizes the 
Pfaffian MR FQH state~\cite{Moore91:MR} in a half-filled Landau 
level~\cite{Greiter91:MR}. We therefore expect an analogous FCI phase 
appearing at half filling. The MR FQH state has fractional 
excitations governed by the generalized Pauli principle of having no more than 
2 particles in 4 consecutive Landau level orbitals and the fermionic 
statistics of no more than one particle allowed per 
orbital~\cite{Bernevig08:Jack,Bernevig08:Jack2}. 
This enables a direct extension of the $(1,3)$-admissible FCI counting of 
having no more than 1 particle in 3 consecutive orbitals~\cite{Regnault11:FCI} 
to a $(2,4)$-admissible counting~\cite{Bernevig11:Counting} for the possible 
MR FCI state.

\begin{figure}[]
\includegraphics{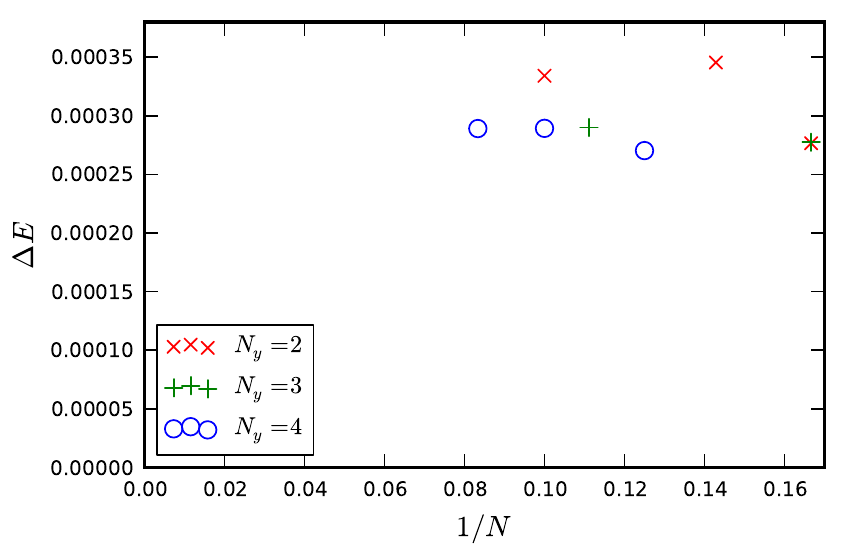}
\caption{\label{fig:kagome-3b-scaling} Energy gap $\Delta E$ of the Kagome lattice 
model with three-body interactions for different system sizes and aspect ratios. 
In each case, $N_x=2N/N_y$.}
\end{figure}

\begin{figure}[]
\includegraphics{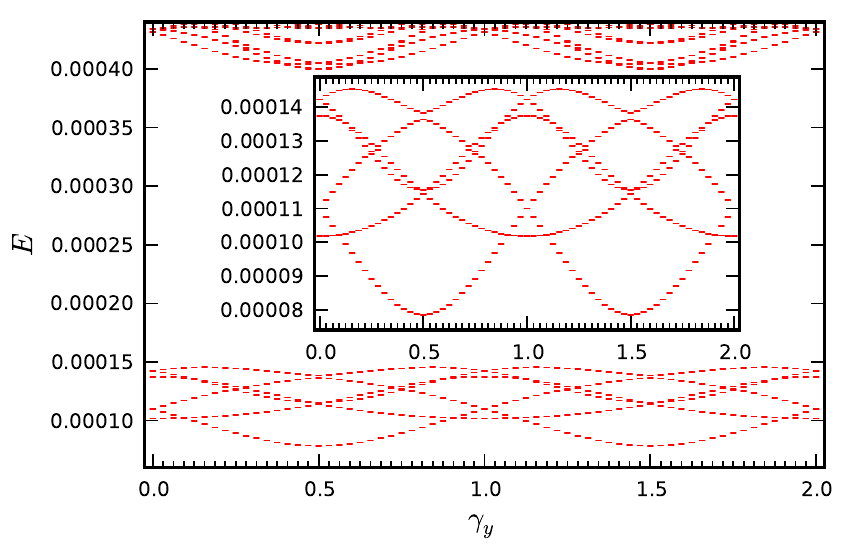}
\caption{\label{fig:kagome-1054twisted} Spectral flow of the low-lying
states of the Kagome lattice model with three-body interaction of $N=10$ 
particles on the
$N_x\times N_y=5\times 4$ lattice upon flux insertion along the $y$ direction. 
$\gamma_y$ counts the number of fluxes inserted. The 6-fold ground states 
flow into each other, and do not mix with higher states during flux insertion. 
After insertion of every 2 full fluxes, the 6-fold states return to the 
original configuration (inset).}
\end{figure}

We diagonalize the interacting Hamiltonian in the flattened lowest band at 
filling $1/2$. We show the energy spectrum of $N=8,10,12$ particles on the
$N_x\times N_y=\frac{N}{2}\times 4$ lattice in 
Fig.~\ref{fig:kagome-844,1054,1264}.
In the three cases, a 6-fold degenerate 
ground state is seen at total momenta $\{6\times(0,0)\}$, 
$\{(0,0),2\times(0,1),(0,2),2\times(0,3)\}$, $\{2\times(0,0),4\times(0,2)\}$, 
respectively. This agrees exactly with the $(2,4)$-admissible counting 
proposed above. 
As shown in Fig.~\ref{fig:kagome-3b-scaling}, the energy gap $\Delta E$ 
remains open and scales to a finite value in the limit of $N\rightarrow\infty$ 
with $N_x/N_y$ finite. 
The six degenerate ground states exhibit spectral flow upon flux 
insertion. As shown in Fig.~\ref{fig:kagome-1054twisted}, the period of the 
spectral flow is 2 fluxes, and therefore the ground state has Hall conductance 
$\sigma_{xy}=1/2$.

\begin{figure}[]
\includegraphics{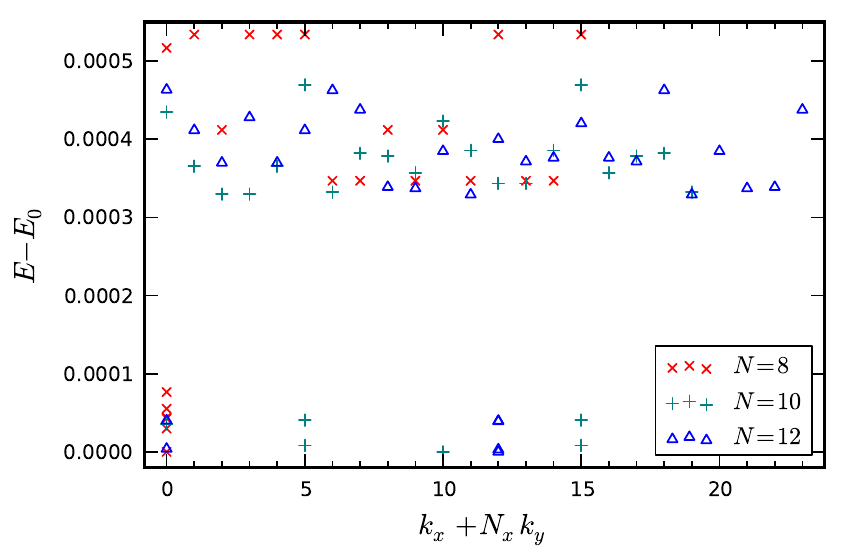}
\caption{\label{fig:kagome-844,1054,1264} Low energy spectrum of the Kagome 
lattice model with three-body interactions of $N=8$ (marked by red crosses), $N=10$ 
(green plus signs), and $N=12$ (blue triangles) particles on the
$N_x\times N_y=\frac{N}{2}\times 4$ lattice, with energies shifted by $E_0$, 
the lowest energy for each system size.
We only show the lowest excited level in each momentum sector in addition to 
the 6-fold ground state. 
}
\end{figure}

\begin{figure}[]
\includegraphics{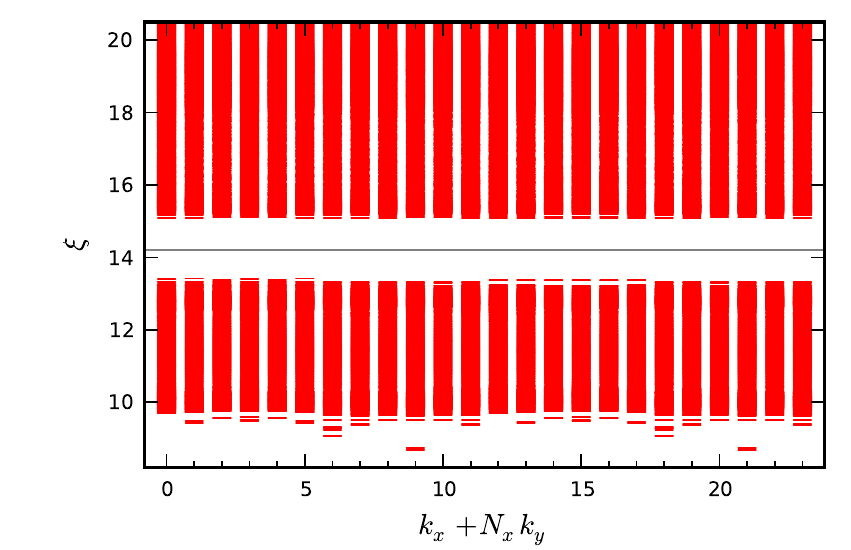}
\caption{\label{fig:kagome-1264pes} Particle entanglement spectrum of the 
ground state of the Kagome lattice model with three-body interactions of $N=12$ 
particles on the $N_x\times N_y=6\times 4$ lattice, with $N_B=6$ particles traced 
out. The number of states below the gray line is $2910$ in momentum sectors 
with $k_x=1,3$ and $k_y=0,2$, $2912$ in sectors $(0,0)$ and $(0,2)$, $2940$ in 
sectors with $k_x=5$ and sectors with $k_x=2,4$ and $k_y=2,4$, and $2944$ in 
all the other sectors, in agreement with the $(2,4)$-admissible counting rule. 
The width of the entanglement gap is $\Delta\xi=1.66$.}
\end{figure}

We probe the quasihole excitations by using the particle entanglement spectrum 
of the ground state. In Fig.~\ref{fig:kagome-1264pes}, we observe a clear and
large gap in the entanglement spectrum, and the counting of the 
entanglement energy levels below the gap again matches in each momentum sector the 
$(2,4)$-admissible counting as predicted in Ref.~\onlinecite{Bernevig11:Counting}. 
The width of the entanglement gap is 
$\Delta\xi=1.66$. This indicates that the inter-electron correlations that obey 
the $(2,4)$ generalized Pauli principle, i.e. the pairing of electrons, is 
5 times stronger than any other kind of correlations.
Therefore, we conclude that the ground state of the Kagome lattice model with 
the three-body NN interactions at half filling is indeed a MR FQH state. 

Based on the results at filling $1/3$ and $1/2$, we make the conjecture that 
\emph{under desirable conditions}, a short-range $(n+1)$-body interaction 
could stabilize $\mathbb{Z}_n$ RR parafermion FQH 
state~\cite{Read99:RR} in a Chern band at filling $n/(n+2)$.  
Further work is needed to confirm the cases $n\geq 3$. The $n=3$ case has 
recently been reported in Ref.~\onlinecite{Bernevig11:Counting}.

We have also looked for a MR phase in the Haldane and two-orbital models using 
three-body interactions. Unfortunately, these two models do not seem to 
exhibit such a phase. This appears to be consistent with the more fragile 
Laughlin-like phase in these models.

\section{Ruby Lattice Model}

\begin{figure}[]
\centering
\includegraphics[]{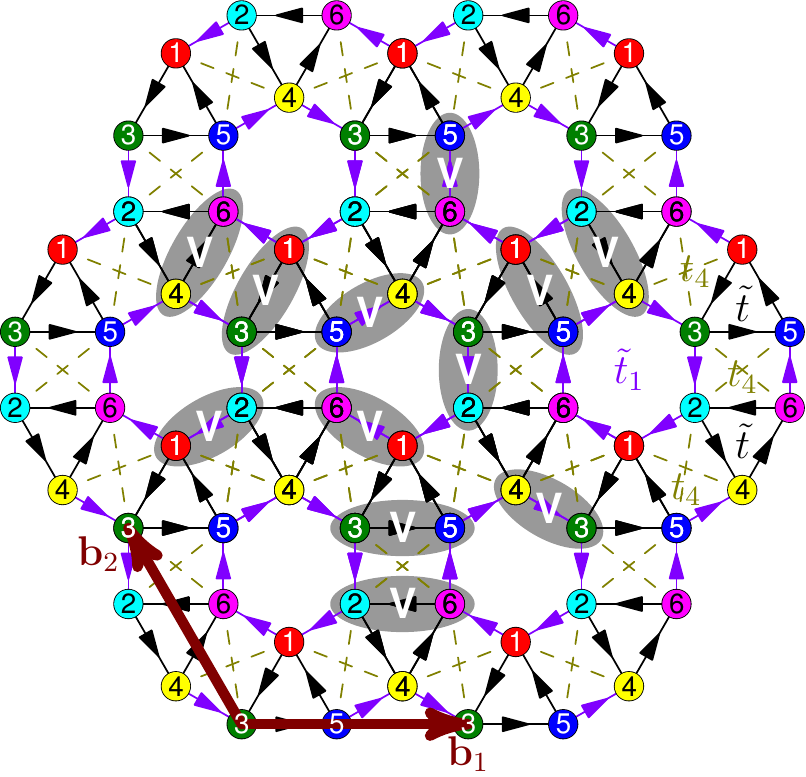}%lbrt
\caption{\label{fig:ruby} The Ruby lattice model. The six sublattices 1 to 6 
are colored respectively in red, cyan, green, yellow, blue and magenta. 
The lattice translation vectors are $\mathbf{b}_1$, $\mathbf{b}_2$.
The complex hopping parameters between NN are $\tilde{t}=t_r + i t_i$ for sites 
having the same parity (black arrows) and $\tilde{t}_1=t_{1r} + i t_{1i}$ for 
sites having opposite parity (purple arrows), both in the direction of the 
arrows. The real hopping parameter on the diagonal of the square (olive dashed 
lines) is given by $t_4$.
The density-density repulsion between NN is depicted by gray 
ellipses.}
\end{figure}

The last Chern insulator we have analyzed is on a ruby lattice.
Hu et al.~\cite{Hu11:Ruby} have considered a two-dimensional 
ruby lattice with strong spin-orbit coupling. The simplified spin-polarized 
version of this model was shown to provide a Chern insulator with an extremely 
flat lowest band.

As depicted in Fig.~\ref{fig:ruby}, the ruby lattice is spanned by the 
translation vectors $\mathbf{b}_1$ and $\mathbf{b}_2$ and it is made of six 
sublattices, denoted from 1 to 6. After a Fourier transform and a trivial 
gauge transform, the single-particle Hamiltonian can be cast in Bloch form as 
\begin{equation}
H=\sum_{{\mathbf k}} \sum_{i,j=1}^{6} \Psi^{\dagger}_{{\mathbf k},i}
h_{i,j}\left({\mathbf k}\right)\Psi_{{\mathbf k},j}.
\end{equation}
Here the lattice momentum 
$\mathbf{k}=(\mathbf{k}\cdot \mathbf{b}_1,\mathbf{k}\cdot \mathbf{b}_2)
\equiv (k_x,k_y)$ is summed over the first Brillouin zone, and the $h(\mathbf{k})$ 
matrix is given by (the upper triangle can be obtained from Hermiticity)
\begin{widetext}
\begin{equation}
h(\mathbf{k})=-\left[
\begin{array}{cccccc}
0 & & & & & \\
\tilde{t}^{*}_1 & 0 & & \phantom{loremipsum} & \text{h.c.} & \phantom{loremipsu} \\
\tilde{t} & \tilde{t}^{*}_1 e^{-i(k_x +k_y )} & 0 & & &\\
t_4 \left(1+e^{ik_x }\right) &  \tilde{t} &  \tilde{t}^{*}_1 e^{ik_x } & 0 & & \\
\tilde{t}^{*} & t_4 \left(1+e^{-i(k_x +k_y )}\right) & \tilde{t} &\tilde{t}^{*}_1 & 0 & \\
\tilde{t}_1 e^{ik_x } & \tilde{t}^{*} & t_4 \left(e^{ik_x }+e^{i(k_x +k_y )}\right) & \tilde{t} & \tilde{t}^{*}_1 e^{i(k_x +k_y )}& 0
\end{array}
\right].
\end{equation}
\end{widetext}

We adopt the parameter values suggested in the original 
article~\cite{Hu11:Ruby}, namely 
$(t_{i},t_{1r},t_{1i},t_4)=(1.2,-1.2,2.6,-1.2)t_{r}$.
For these parameters, the lowest band of the problem is gapped, and has unit 
Chern number~\cite{Hu11:Ruby}.

\begin{figure}[]
\includegraphics{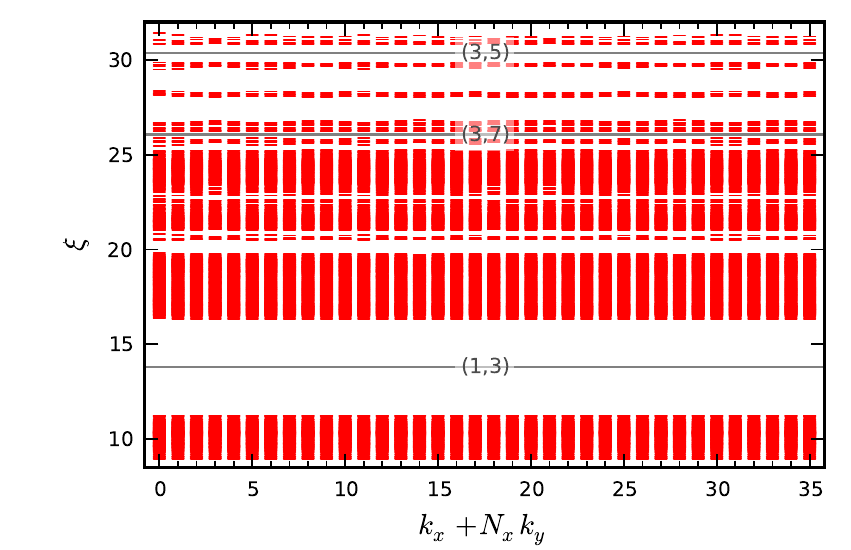}
\caption{\label{fig:ruby-1266pes} Particle entanglement spectrum of the ground 
state of the ruby lattice model of $N=12$ particles on the 
$N_x\times N_y=6\times 6$ lattice, with $N_B=8$ particles traced out. The 
number of states below each of 
the three gray lines is $741,1629,1645$ in momentum sectors where both $k_x$ 
and $k_y$ are even and $728,1612,1628$ elsewhere.
These numbers are in perfect accordance with the $(1,3)$-, $(3,7)$-, and 
$(3,5)$-admissible counting, respectively. 
The width of the $(1,3)$ entanglement gap is $\Delta \xi=5.07$.
The counting of levels below the unmarked gaps do not correspond 
to any $(k,r)$-admissible rule.
}
\end{figure}

Again, we flatten the Bloch bands using projectors, and we add density-density 
repulsion of unit strength between nearest neighbors. There are 12 type of 
terms as depicted by the gray ellipses in Fig.~\ref{fig:ruby}.
We diagonalize the interacting Hamiltonian in the flattened lowest band at 
filling $1/3$ with up to $N=12$ particles, and find results quite similar to 
the previous three models. Namely, we find a 3-fold degenerate ground state 
in the momentum sectors predicted by the $(1,3)$ counting principle. The 
degenerate ground state is separated from the excitations by a finite energy gap. 
Finite-size scaling indicates that the gap remains open in the thermodynamic 
limit.
Twisting boundary drives spectral flow within the ground-state manifold and 
the flow has a period of 3 fluxes. 
Hence, the system has Hall conductance $\sigma_{xy}=1/3$. 
We calculate the entanglement spectra for various system sizes and find an 
entanglement gap with the $(1,3)$-admissible counting in each case. 
As an example, we show in Fig.~\ref{fig:ruby-1266pes} the entanglement 
spectrum of the ground state of $N=12$ particles. The entanglement gap 
$\Delta \xi=5.07$ is comparable to, although slightly smaller than the 
entanglement gap $\Delta \xi=5.53$ of the Kagome lattice model with only NN 
hoppings at the same system size.
This pronounced sign of exclusion statistics counting suggests that the system has 
fractional excitations similar to the Laughlin quasiholes. 
We do however stress that we have \emph{not} computed the statistics of the 
quasiholes, as this would require braiding operations likely to be plagued by 
finite-size effects on the lattices we can reach by computers. We 
conclude that the ground state of the ruby lattice model with two-body NN 
repulsions is a FQH Laughlin state.

We have also checked the effect of three-body NN repulsions at half filling. 
There are a total of 14 terms, corresponding to 2 equilateral triangles 
and 12 right triangles embedded in rectangles.
Similar to the case of the Kagome lattice model detailed in 
Section~\ref{sec:kagome-half-filling}, we find a 
robust MR state here. The state has the hallmark 6-fold degenerate gapped 
ground state at the correct momenta. We also find a large entanglement gap
corresponding to fractional excitations governed by the $(2,4)$ generalized 
Pauli principle. 

\section{Structures in entanglement spectrum}

As discussed in previous sections, the entanglement spectrum of the ground 
state of a short-range $(m+1)$-body interaction has a gap corresponding to the 
$(m,m+2)$-admissible counting, for $m=1,2$. This gap measures the prominence 
of $m$-particle clustering in the ground state. The $(m+1)$-body interactions 
could possibly generate clusters of other sizes as well.
For example, in the FQH effect in the continuum, MR states can be obtained with just 
two-body potentials. In that case, there could be additional gaps in the 
entanglement spectrum, and the counting of levels below these gaps could 
conform to other generalized Pauli principles.

The appearance of a $(n,n+r)$-admissible 
counting reveals the existence of a specific clustering pattern in the ground 
state. This pattern is also present in the model fermionic FQH state at 
filling $\nu=n/(n+r)$ described by the $(n,r)$ Jack polynomials multiplied by 
a Vandermonde determinant~\cite{Bernevig08:Jack,Bernevig08:Jack2}. 
For example, a $(n,n+2)$-admissible counting corresponds to the $\mathbb{Z}_n$ 
Read-Rezayi FQH state~\cite{Read99:RR}, and a $(2,5)$-admissible counting 
corresponds to the `Gaffnian' FQH wave function~\cite{Simon07:Gaffnian}. The model wave 
functions of such FQH states have characteristic zeros when a cluster of $n+1$ 
particles forms even after removing the ``trivial'' zeros provided by the 
fermionic statistics. This reflects specific $(n+1)$-body correlations in the system.
Therefore, the presence of a gap with a $(n,n+r)$ admissible counting in the 
FCI entanglement spectrum signals the presence of clustering correlations 
similar to a specific model FQH state, and implies the presence of stable 
$(n+1)$-body correlations in the system.

\begin{figure}[]
\includegraphics{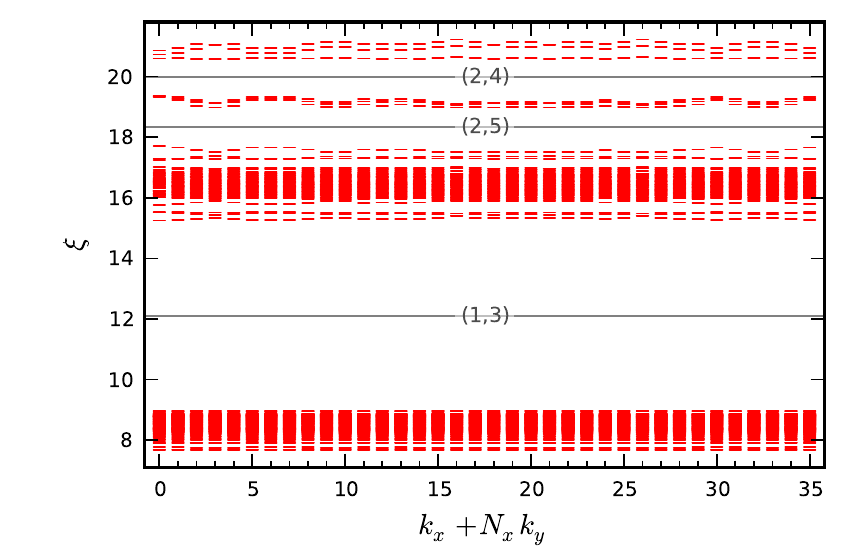}
\caption{\label{fig:kagome-1266pes-3} Particle entanglement spectrum of the ground 
state of the Kagome lattice model of $N=12$ particles on the 
$N_x\times N_y=6\times 6$ 
lattice, with $N_B=9$ particles traced out. The number of states below each of 
the three gray lines is $138,195,198$ in momentum sectors where $k_x,k_y\in\{0,3\}$ 
and $135,192,195$ elsewhere. These numbers are in perfect accordance with the 
$(1,3)$-, $(2,5)$-, and $(2,4)$-admissible counting, respectively.
}
\end{figure}

We check comprehensively all the four models studied in this paper as well 
as the checkerboard lattice model~\cite{Sun11:Flatband,Regnault11:FCI}, at 
various system sizes up to $N=12$ particles. In most cases, we find extra
entanglement gaps other than the one above the states of $(m,m+2)$-admissible 
counting in the ground state of an $(m+1)$-body interaction, and the counting 
of entanglement energy levels below \emph{some} of these 
gaps matches with an $(n,n+r)$-admissible counting in each momentum 
sector. For all the systems we look at, we usually find 
at least one such extra gap that matches perfectly with a particular $(n,n+r)$-admissible 
rule. For example, in the Kagome lattice model with two-body interaction, we 
find two extra entanglement gaps in the ground state 
(Fig.~\ref{fig:kagome-1266pes-3}), with the counting of levels below them 
given \emph{exactly} (in each momentum sector) by the $(2,4)$- and $(2,5)$-admissible 
counting and the folding based on the FQH-FCI 
mapping~\cite{Bernevig11:Counting}. The former case corresponds 
to the MR state while the latter corresponds to the Gaffnian wave function.
We find the appearance of the counting of non-Abelian statistics in the 
Laughlin-like ground state of two-body interactions quite intriguing.

We do not have a complete understanding of the full gap structure 
at the moment. In the earlier example of the ruby lattice model 
shown in Fig.~\ref{fig:ruby-1266pes}, we find a few extra entanglement gaps. 
Two of them have counting below the gap given \emph{exactly} (in each momentum 
sector) by $(3,5)$- and $(3,7)$-admissible rules~\cite{Bernevig11:Counting} 
respectively, the former of which corresponds to a $\mathbb{Z}_3$ Read-Rezayi 
state. There are other entanglement gaps in the spectrum that cannot be 
explained by any $(n,n+r)$-admissible rule.

All of such extra $(n,n+r)$ entanglement gaps that we observe have 
$n+1=N_A$, where $N_A$ is the number of particles left in the system when 
evaluating the entanglement spectrum. Tracing out part of the system enhances 
the correlation between the remaining $N_A$ particles. One can imagine that such 
enhancement is more significant for the correlation between the $N_A$ 
particles as a whole than the correlation within a fraction of the $N_A$ 
particles. In most cases, we observe such $(N_A-1,N_A-1+r)$ entanglement gaps 
for almost every value of $N_A$. 
Our examples show that a ground-state wave function at $\nu=1/3$ can give us 
information about possible non-Abelian correlations.

This suggests a feature of the FCI: the $(n+1)$-body 
interacting Hamiltonian could not only generate strong $(n+1)$-body 
correlations, but also higher-body correlations. This gives hope 
for the possible realization of other members of the Read-Rezayi series with
two-body potentials. It is very intriguing that entanglement in the ground 
state of an $(n+1)$-body interacting Hamiltonian at filling $\nu=n/(n+2)$ 
contains clear clues that some other FQH states could possibly be stabilized 
by some \emph{other} interaction at some \emph{other} filling.

\section{Discussion and Conclusion}

In this paper, we have studied four examples of flat-band Chern 
insulators. If we include the previous 
articles~\cite{Sheng11:FCI,Neupert11:FCI,Wang11:FCI-Boson,Regnault11:FCI} on 
the checkerboard lattice, we now have five cases which can be used to compare 
the universal physics. 

First, we have showed that the existence of FCI does not 
require a special lattice. Under proper conditions, FCI phases could (at 
least) emerge from lattice models with 2, 3, or 6 sites per unit cell. There 
is no restriction on the type of Bravais lattice either.

Second, we find concrete evidence that the stability of the FCI phase is not 
guaranteed by flat band dispersion. A large enough gap to bandwidth ratio is 
necessary for the existence of an insulating state at a fractional filling.
Our calculations performed in the flat-band limit do not 
capture this effect. However, our results reveal that the single-particle 
eigenstates play a critical role in the formation of a FCI state as well; a 
badly chosen parameter set could destroy the topological phase even when the 
band dispersion is perfectly flat.
The example of the Kagome model is especially instructive. While the addition 
of NNN hopping terms could make the band flatter, it actually 
deteriorates the FCI phase. Thus the search of a realistic 
FCI that would include the effect of the band dispersion relation, should not 
focus only on the flat-band criterion.

Third, the development of a FCI state is often dependent on suppressed fluctuations 
in the Berry curvature. This agrees qualitatively with the picture derived 
from the algebra of projected density operators. It should be noted, however, 
that even the suppressed curvature fluctuations could actually be considerably strong.

Fourth, out of the five models that we have checked, the Kagome lattice model 
with only NN hoppings has the most robust FCI phase. 
The ruby lattice model comes close, but has three more tunable parameters.

In conclusion, we have established the existence of a FCI phase in four
distinct flat-band lattice models at filling $\nu=1/3$ with short-range 
repulsive density-density interactions. The FCI phase is identified by an 
incompressible ground state with Hall conductance $\sigma_{xy}=1/3$ and 
excitations obeying the same counting as Laughlin quasiholes. 
We have observed that in some cases, band structures that favor the emergence 
of the FCI phase have less fluctuating Berry curvature.
In the presence of short-range three-body repulsive interactions,
we have also found another FCI phase reminiscent of the Pfaffian Moore-Read FQH 
state at half filling of the Kagome lattice model and the ruby lattice model, 
in addition to the previously known example of the Moore-Read phase in the 
checkerboard lattice model. This more exotic 
phase has an incompressible ground state with Hall conductance 
$\sigma_{xy}=1/2$ and excitations exhibiting the counting of Moore-Read 
quasiholes. On the technical side, we have demonstrated from 
various angles the power of the entanglement spectrum as a sensitive and reliable 
probe of topological order, when no model wave functions are available. We 
have also discussed interesting structures in the entanglement spectrum and 
their implication for the possible stable existence of other FQH states at 
zero magnetic field. Future directions for this line of research include 
strongly interacting Chern insulators with Chern number higher than one, such 
as the dice model\cite{Wang11:Dice}.

\section{Acknowledgements}
BAB wishes to thank Z.~Papic, F.D.M.~Haldane, T.L.~Hughes, S.L.~Sondhi, and 
S.A.~Parameswaran for useful discussions. NR acknowledges fruitful discussions 
with G.~Moeller.  BAB was supported by Princeton Startup Funds, NSF CAREER 
DMR-095242, ONR - N00014-11-1-0635, DARPA - N66001-11-1-4110 and NSF-MRSEC 
DMR-0819860 at Princeton University. YLW was supported by NSF CAREER 
DMR-095242. BAB thanks Technion, Israel, and Ecole Normale Superieure, Paris, 
for generous hosting during the stages of this work. 

\bibliography{cond-mat}

\begin{thebibliography}{48}
\expandafter\ifx\csname natexlab\endcsname\relax\def\natexlab#1{#1}\fi
\expandafter\ifx\csname bibnamefont\endcsname\relax
  \def\bibnamefont#1{#1}\fi
\expandafter\ifx\csname bibfnamefont\endcsname\relax
  \def\bibfnamefont#1{#1}\fi
\expandafter\ifx\csname citenamefont\endcsname\relax
  \def\citenamefont#1{#1}\fi
\expandafter\ifx\csname url\endcsname\relax
  \def\url#1{\texttt{#1}}\fi
\expandafter\ifx\csname urlprefix\endcsname\relax\def\urlprefix{URL }\fi
\providecommand{\bibinfo}[2]{#2}
\providecommand{\eprint}[2][]{\url{#2}}

\bibitem[{\citenamefont{Haldane}(1988)}]{Haldane88:Honeycomb}
\bibinfo{author}{\bibfnamefont{F.~D.~M.} \bibnamefont{Haldane}},
  \bibinfo{journal}{Physical Review Letters} \textbf{\bibinfo{volume}{61}},
  \bibinfo{pages}{2015} (\bibinfo{year}{1988}).

\bibitem[{\citenamefont{Kane and Mele}(2005{\natexlab{a}})}]{Kane05:KM}
\bibinfo{author}{\bibfnamefont{C.~L.} \bibnamefont{Kane}} \bibnamefont{and}
  \bibinfo{author}{\bibfnamefont{E.~J.} \bibnamefont{Mele}},
  \bibinfo{journal}{Physical Review Letters} \textbf{\bibinfo{volume}{95}},
  \bibinfo{pages}{146802} (\bibinfo{year}{2005}{\natexlab{a}}).

\bibitem[{\citenamefont{Kane and Mele}(2005{\natexlab{b}})}]{Kane05:KM2}
\bibinfo{author}{\bibfnamefont{C.~L.} \bibnamefont{Kane}} \bibnamefont{and}
  \bibinfo{author}{\bibfnamefont{E.~J.} \bibnamefont{Mele}},
  \bibinfo{journal}{Physical Review Letters} \textbf{\bibinfo{volume}{95}},
  \bibinfo{pages}{226801} (\bibinfo{year}{2005}{\natexlab{b}}).

\bibitem[{\citenamefont{Bernevig and Zhang}(2006)}]{Bernevig06:QSH}
\bibinfo{author}{\bibfnamefont{B.~A.} \bibnamefont{Bernevig}} \bibnamefont{and}
  \bibinfo{author}{\bibfnamefont{S.-C.} \bibnamefont{Zhang}},
  \bibinfo{journal}{Physical Review Letters} \textbf{\bibinfo{volume}{96}},
  \bibinfo{pages}{106802} (\bibinfo{year}{2006}).

\bibitem[{\citenamefont{Bernevig et~al.}(2006)\citenamefont{Bernevig, Hughes,
  and Zhang}}]{Bernevig06:BHZ}
\bibinfo{author}{\bibfnamefont{B.~A.} \bibnamefont{Bernevig}},
  \bibinfo{author}{\bibfnamefont{T.~L.} \bibnamefont{Hughes}},
  \bibnamefont{and} \bibinfo{author}{\bibfnamefont{S.-C.} \bibnamefont{Zhang}},
  \bibinfo{journal}{Science} \textbf{\bibinfo{volume}{314}},
  \bibinfo{pages}{1757} (\bibinfo{year}{2006}).

\bibitem[{\citenamefont{K{\"{o}}nig et~al.}(2007)\citenamefont{K{\"{o}}nig,
  Wiedmann, Br{\"{u}}ne, Roth, Buhmann, Molenkamp, Qi, and
  Zhang}}]{Konig07:QSH}
\bibinfo{author}{\bibfnamefont{M.}~\bibnamefont{K{\"{o}}nig}},
  \bibinfo{author}{\bibfnamefont{S.}~\bibnamefont{Wiedmann}},
  \bibinfo{author}{\bibfnamefont{C.}~\bibnamefont{Br{\"{u}}ne}},
  \bibinfo{author}{\bibfnamefont{A.}~\bibnamefont{Roth}},
  \bibinfo{author}{\bibfnamefont{H.}~\bibnamefont{Buhmann}},
  \bibinfo{author}{\bibfnamefont{L.~W.} \bibnamefont{Molenkamp}},
  \bibinfo{author}{\bibfnamefont{X.-L.} \bibnamefont{Qi}}, \bibnamefont{and}
  \bibinfo{author}{\bibfnamefont{S.-C.} \bibnamefont{Zhang}},
  \bibinfo{journal}{Science} \textbf{\bibinfo{volume}{318}},
  \bibinfo{pages}{766} (\bibinfo{year}{2007}).

\bibitem[{\citenamefont{Hsieh et~al.}(2008)\citenamefont{Hsieh, Qian, Wray,
  Xia, Hor, Cava, and Hasan}}]{Hsieh08:BiSb}
\bibinfo{author}{\bibfnamefont{D.}~\bibnamefont{Hsieh}},
  \bibinfo{author}{\bibfnamefont{D.}~\bibnamefont{Qian}},
  \bibinfo{author}{\bibfnamefont{L.}~\bibnamefont{Wray}},
  \bibinfo{author}{\bibfnamefont{Y.}~\bibnamefont{Xia}},
  \bibinfo{author}{\bibfnamefont{Y.~S.} \bibnamefont{Hor}},
  \bibinfo{author}{\bibfnamefont{R.~J.} \bibnamefont{Cava}}, \bibnamefont{and}
  \bibinfo{author}{\bibfnamefont{M.~Z.} \bibnamefont{Hasan}},
  \bibinfo{journal}{Nature} \textbf{\bibinfo{volume}{452}},
  \bibinfo{pages}{970} (\bibinfo{year}{2008}).

\bibitem[{\citenamefont{Hasan and Kane}(2010)}]{Hasan10:RMP}
\bibinfo{author}{\bibfnamefont{M.}~\bibnamefont{Hasan}} \bibnamefont{and}
  \bibinfo{author}{\bibfnamefont{C.}~\bibnamefont{Kane}},
  \bibinfo{journal}{Reviews of Modern Physics} \textbf{\bibinfo{volume}{82}},
  \bibinfo{pages}{3045} (\bibinfo{year}{2010}).

\bibitem[{\citenamefont{Hasan and Moore}(2011)}]{Hasan11:AnnRev}
\bibinfo{author}{\bibfnamefont{M.~Z.} \bibnamefont{Hasan}} \bibnamefont{and}
  \bibinfo{author}{\bibfnamefont{J.~E.} \bibnamefont{Moore}},
  \bibinfo{journal}{Annual Review of Condensed Matter Physics}
  \textbf{\bibinfo{volume}{2}}, \bibinfo{pages}{55} (\bibinfo{year}{2011}).

\bibitem[{\citenamefont{Schnyder et~al.}(2008)\citenamefont{Schnyder, Ryu,
  Furusaki, and Ludwig}}]{Schnyder08:Table}
\bibinfo{author}{\bibfnamefont{A.}~\bibnamefont{Schnyder}},
  \bibinfo{author}{\bibfnamefont{S.}~\bibnamefont{Ryu}},
  \bibinfo{author}{\bibfnamefont{A.}~\bibnamefont{Furusaki}}, \bibnamefont{and}
  \bibinfo{author}{\bibfnamefont{A.}~\bibnamefont{Ludwig}},
  \bibinfo{journal}{Physical Review B} \textbf{\bibinfo{volume}{78}},
  \bibinfo{pages}{195125} (\bibinfo{year}{2008}).

\bibitem[{\citenamefont{Qi et~al.}(2008)\citenamefont{Qi, Hughes, and
  Zhang}}]{Qi08:Table}
\bibinfo{author}{\bibfnamefont{X.-L.} \bibnamefont{Qi}},
  \bibinfo{author}{\bibfnamefont{T.~L.} \bibnamefont{Hughes}},
  \bibnamefont{and} \bibinfo{author}{\bibfnamefont{S.-C.} \bibnamefont{Zhang}},
  \bibinfo{journal}{Physical Review B} \textbf{\bibinfo{volume}{78}},
  \bibinfo{pages}{195424} (\bibinfo{year}{2008}).

\bibitem[{\citenamefont{Kitaev}(2009)}]{Kitaev09:Table}
\bibinfo{author}{\bibfnamefont{A.}~\bibnamefont{Kitaev}}, \bibinfo{journal}{AIP
  Conference Proceedings} \textbf{\bibinfo{volume}{1134}}, \bibinfo{pages}{22}
  (\bibinfo{year}{2009}).

\bibitem[{\citenamefont{Neupert
  et~al.}(2011{\natexlab{a}})\citenamefont{Neupert, Santos, Chamon, and
  Mudry}}]{Neupert11:FCI}
\bibinfo{author}{\bibfnamefont{T.}~\bibnamefont{Neupert}},
  \bibinfo{author}{\bibfnamefont{L.}~\bibnamefont{Santos}},
  \bibinfo{author}{\bibfnamefont{C.}~\bibnamefont{Chamon}}, \bibnamefont{and}
  \bibinfo{author}{\bibfnamefont{C.}~\bibnamefont{Mudry}},
  \bibinfo{journal}{Physical Review Letters} \textbf{\bibinfo{volume}{106}},
  \bibinfo{pages}{236804} (\bibinfo{year}{2011}{\natexlab{a}}).

\bibitem[{\citenamefont{Sheng et~al.}(2011)\citenamefont{Sheng, Gu, Sun, and
  Sheng}}]{Sheng11:FCI}
\bibinfo{author}{\bibfnamefont{D.~N.} \bibnamefont{Sheng}},
  \bibinfo{author}{\bibfnamefont{Z.-C.} \bibnamefont{Gu}},
  \bibinfo{author}{\bibfnamefont{K.}~\bibnamefont{Sun}}, \bibnamefont{and}
  \bibinfo{author}{\bibfnamefont{L.}~\bibnamefont{Sheng}},
  \bibinfo{journal}{Nature Communications} \textbf{\bibinfo{volume}{2}},
  \bibinfo{pages}{389} (\bibinfo{year}{2011}).

\bibitem[{\citenamefont{Regnault and Bernevig}(2011)}]{Regnault11:FCI}
\bibinfo{author}{\bibfnamefont{N.}~\bibnamefont{Regnault}} \bibnamefont{and}
  \bibinfo{author}{\bibfnamefont{B.~A.} \bibnamefont{Bernevig}},
  \bibinfo{journal}{ArXiv e-prints}  (\bibinfo{year}{2011}),
  \eprint{1105.4867}.

\bibitem[{\citenamefont{Venderbos et~al.}(2011)\citenamefont{Venderbos,
  Kourtis, van~den Brink, and Daghofer}}]{Venderbos11:t2g}
\bibinfo{author}{\bibfnamefont{J.~W.~F.} \bibnamefont{Venderbos}},
  \bibinfo{author}{\bibfnamefont{S.}~\bibnamefont{Kourtis}},
  \bibinfo{author}{\bibfnamefont{J.}~\bibnamefont{van~den Brink}},
  \bibnamefont{and} \bibinfo{author}{\bibfnamefont{M.}~\bibnamefont{Daghofer}},
  \bibinfo{journal}{ArXiv e-prints}  (\bibinfo{year}{2011}),
  \eprint{1109.5955}.

\bibitem[{\citenamefont{Wang et~al.}(2011{\natexlab{a}})\citenamefont{Wang, Gu,
  Gong, and Sheng}}]{Wang11:FCI-Boson}
\bibinfo{author}{\bibfnamefont{Y.-F.} \bibnamefont{Wang}},
  \bibinfo{author}{\bibfnamefont{Z.-C.} \bibnamefont{Gu}},
  \bibinfo{author}{\bibfnamefont{C.-D.} \bibnamefont{Gong}}, \bibnamefont{and}
  \bibinfo{author}{\bibfnamefont{D.}~\bibnamefont{Sheng}},
  \bibinfo{journal}{Physical Review Letters} \textbf{\bibinfo{volume}{107}},
  \bibinfo{pages}{146803} (\bibinfo{year}{2011}{\natexlab{a}}).

\bibitem[{\citenamefont{Wang et~al.}(2011{\natexlab{b}})\citenamefont{Wang,
  Yao, Gu, Gong, and Sheng}}]{Wang11:MR}
\bibinfo{author}{\bibfnamefont{Y.-F.} \bibnamefont{Wang}},
  \bibinfo{author}{\bibfnamefont{H.}~\bibnamefont{Yao}},
  \bibinfo{author}{\bibfnamefont{Z.-C.} \bibnamefont{Gu}},
  \bibinfo{author}{\bibfnamefont{C.-D.} \bibnamefont{Gong}}, \bibnamefont{and}
  \bibinfo{author}{\bibfnamefont{D.~N.} \bibnamefont{Sheng}},
  \bibinfo{journal}{ArXiv e-prints}  (\bibinfo{year}{2011}{\natexlab{b}}),
  \eprint{1110.4980}.

\bibitem[{\citenamefont{Neupert
  et~al.}(2011{\natexlab{b}})\citenamefont{Neupert, Santos, Ryu, Chamon, and
  Mudry}}]{Neupert11:Z2}
\bibinfo{author}{\bibfnamefont{T.}~\bibnamefont{Neupert}},
  \bibinfo{author}{\bibfnamefont{L.}~\bibnamefont{Santos}},
  \bibinfo{author}{\bibfnamefont{S.}~\bibnamefont{Ryu}},
  \bibinfo{author}{\bibfnamefont{C.}~\bibnamefont{Chamon}}, \bibnamefont{and}
  \bibinfo{author}{\bibfnamefont{C.}~\bibnamefont{Mudry}},
  \bibinfo{journal}{Physical Review B} \textbf{\bibinfo{volume}{84}},
  \bibinfo{pages}{165107} (\bibinfo{year}{2011}{\natexlab{b}}).

\bibitem[{\citenamefont{Santos et~al.}(2011)\citenamefont{Santos, Neupert, Ryu,
  Chamon, and Mudry}}]{Santos11:BF}
\bibinfo{author}{\bibfnamefont{L.}~\bibnamefont{Santos}},
  \bibinfo{author}{\bibfnamefont{T.}~\bibnamefont{Neupert}},
  \bibinfo{author}{\bibfnamefont{S.}~\bibnamefont{Ryu}},
  \bibinfo{author}{\bibfnamefont{C.}~\bibnamefont{Chamon}}, \bibnamefont{and}
  \bibinfo{author}{\bibfnamefont{C.}~\bibnamefont{Mudry}},
  \bibinfo{journal}{ArXiv e-prints}  (\bibinfo{year}{2011}),
  \eprint{1108.2440}.

\bibitem[{\citenamefont{Neupert
  et~al.}(2011{\natexlab{c}})\citenamefont{Neupert, Santos, Ryu, Chamon, and
  Mudry}}]{Neupert11:Hubbard}
\bibinfo{author}{\bibfnamefont{T.}~\bibnamefont{Neupert}},
  \bibinfo{author}{\bibfnamefont{L.}~\bibnamefont{Santos}},
  \bibinfo{author}{\bibfnamefont{S.}~\bibnamefont{Ryu}},
  \bibinfo{author}{\bibfnamefont{C.}~\bibnamefont{Chamon}}, \bibnamefont{and}
  \bibinfo{author}{\bibfnamefont{C.}~\bibnamefont{Mudry}},
  \bibinfo{journal}{ArXiv e-prints}  (\bibinfo{year}{2011}{\natexlab{c}}),
  \eprint{1110.1296}.

\bibitem[{\citenamefont{Bernevig and Regnault}(2011)}]{Bernevig11:Counting}
\bibinfo{author}{\bibfnamefont{B.~A.} \bibnamefont{Bernevig}} \bibnamefont{and}
  \bibinfo{author}{\bibfnamefont{N.}~\bibnamefont{Regnault}},
  \bibinfo{journal}{ArXiv e-prints}  (\bibinfo{year}{2011}),
  \eprint{1110.4488}.

\bibitem[{\citenamefont{Bernevig and
  Haldane}(2008{\natexlab{a}})}]{Bernevig08:Jack}
\bibinfo{author}{\bibfnamefont{B.~A.} \bibnamefont{Bernevig}} \bibnamefont{and}
  \bibinfo{author}{\bibfnamefont{F.~D.~M.} \bibnamefont{Haldane}},
  \bibinfo{journal}{Physical Review Letters} \textbf{\bibinfo{volume}{100}},
  \bibinfo{pages}{246802} (\bibinfo{year}{2008}{\natexlab{a}}).

\bibitem[{\citenamefont{Bernevig and
  Haldane}(2008{\natexlab{b}})}]{Bernevig08:Jack2}
\bibinfo{author}{\bibfnamefont{B.~A.} \bibnamefont{Bernevig}} \bibnamefont{and}
  \bibinfo{author}{\bibfnamefont{F.~D.~M.} \bibnamefont{Haldane}},
  \bibinfo{journal}{Physical Review Letters} \textbf{\bibinfo{volume}{101}},
  \bibinfo{pages}{246806} (\bibinfo{year}{2008}{\natexlab{b}}).

\bibitem[{\citenamefont{Parameswaran et~al.}(2011)\citenamefont{Parameswaran,
  Roy, and Sondhi}}]{Parameswaran11:W-inf}
\bibinfo{author}{\bibfnamefont{S.~A.} \bibnamefont{Parameswaran}},
  \bibinfo{author}{\bibfnamefont{R.}~\bibnamefont{Roy}}, \bibnamefont{and}
  \bibinfo{author}{\bibfnamefont{S.~L.} \bibnamefont{Sondhi}},
  \bibinfo{journal}{ArXiv e-prints}  (\bibinfo{year}{2011}),
  \eprint{1106.4025}.

\bibitem[{\citenamefont{Murthy and Shankar}(2011)}]{Murthy11:CF}
\bibinfo{author}{\bibfnamefont{G.}~\bibnamefont{Murthy}} \bibnamefont{and}
  \bibinfo{author}{\bibfnamefont{R.}~\bibnamefont{Shankar}},
  \bibinfo{journal}{ArXiv e-prints}  (\bibinfo{year}{2011}),
  \eprint{1108.5501}.

\bibitem[{\citenamefont{Girvin et~al.}(1986)\citenamefont{Girvin, MacDonald,
  and Platzman}}]{Girvin86:GMP}
\bibinfo{author}{\bibfnamefont{S.~M.} \bibnamefont{Girvin}},
  \bibinfo{author}{\bibfnamefont{A.~H.} \bibnamefont{MacDonald}},
  \bibnamefont{and} \bibinfo{author}{\bibfnamefont{P.~M.}
  \bibnamefont{Platzman}}, \bibinfo{journal}{Physical Review B}
  \textbf{\bibinfo{volume}{33}}, \bibinfo{pages}{2481} (\bibinfo{year}{1986}).

\bibitem[{\citenamefont{Qi}(2011)}]{Qi11:Wavefunction}
\bibinfo{author}{\bibfnamefont{X.-L.} \bibnamefont{Qi}},
  \bibinfo{journal}{Physical Review Letters} \textbf{\bibinfo{volume}{107}},
  \bibinfo{pages}{126803} (\bibinfo{year}{2011}).

\bibitem[{\citenamefont{Lu and Ran}(2011)}]{Lu11:Parton}
\bibinfo{author}{\bibfnamefont{Y.~M.} \bibnamefont{Lu}} \bibnamefont{and}
  \bibinfo{author}{\bibfnamefont{Y.}~\bibnamefont{Ran}},
  \bibinfo{journal}{ArXiv e-prints}  (\bibinfo{year}{2011}),
  \eprint{1109.0226}.

\bibitem[{\citenamefont{McGreevy et~al.}(2011)\citenamefont{McGreevy, Swingle,
  and Tran}}]{McGreevy11:Parton}
\bibinfo{author}{\bibfnamefont{J.}~\bibnamefont{McGreevy}},
  \bibinfo{author}{\bibfnamefont{B.}~\bibnamefont{Swingle}}, \bibnamefont{and}
  \bibinfo{author}{\bibfnamefont{K.~A.} \bibnamefont{Tran}},
  \bibinfo{journal}{ArXiv e-prints}  (\bibinfo{year}{2011}),
  \eprint{1109.1569}.

\bibitem[{\citenamefont{Vaezi}(2011)}]{Vaezi11:Parton}
\bibinfo{author}{\bibfnamefont{A.}~\bibnamefont{Vaezi}},
  \bibinfo{journal}{ArXiv e-prints}  (\bibinfo{year}{2011}),
  \eprint{1105.0406}.

\bibitem[{\citenamefont{Tang et~al.}(2011)\citenamefont{Tang, Mei, and
  Wen}}]{Tang11:Kagome}
\bibinfo{author}{\bibfnamefont{E.}~\bibnamefont{Tang}},
  \bibinfo{author}{\bibfnamefont{J.-W.} \bibnamefont{Mei}}, \bibnamefont{and}
  \bibinfo{author}{\bibfnamefont{X.-G.} \bibnamefont{Wen}},
  \bibinfo{journal}{Physical Review Letters} \textbf{\bibinfo{volume}{106}},
  \bibinfo{pages}{236802} (\bibinfo{year}{2011}).

\bibitem[{\citenamefont{Hu et~al.}(2011)\citenamefont{Hu, Kargarian, and
  Fiete}}]{Hu11:Ruby}
\bibinfo{author}{\bibfnamefont{X.}~\bibnamefont{Hu}},
  \bibinfo{author}{\bibfnamefont{M.}~\bibnamefont{Kargarian}},
  \bibnamefont{and} \bibinfo{author}{\bibfnamefont{G.}~\bibnamefont{Fiete}},
  \bibinfo{journal}{Physical Review B} \textbf{\bibinfo{volume}{84}},
  \bibinfo{pages}{155116} (\bibinfo{year}{2011}).

\bibitem[{\citenamefont{Moore and Read}(1991)}]{Moore91:MR}
\bibinfo{author}{\bibfnamefont{G.}~\bibnamefont{Moore}} \bibnamefont{and}
  \bibinfo{author}{\bibfnamefont{N.}~\bibnamefont{Read}},
  \bibinfo{journal}{Nuclear Physics B} \textbf{\bibinfo{volume}{360}},
  \bibinfo{pages}{362} (\bibinfo{year}{1991}).

\bibitem[{\citenamefont{Li and Haldane}(2008)}]{Li08:ES}
\bibinfo{author}{\bibfnamefont{H.}~\bibnamefont{Li}} \bibnamefont{and}
  \bibinfo{author}{\bibfnamefont{F.~D.~M.} \bibnamefont{Haldane}},
  \bibinfo{journal}{Physical Review Letters} \textbf{\bibinfo{volume}{101}},
  \bibinfo{pages}{010504} (\bibinfo{year}{2008}).

\bibitem[{\citenamefont{Sterdyniak et~al.}(2011)\citenamefont{Sterdyniak,
  Regnault, and Bernevig}}]{Sterdyniak11:PES}
\bibinfo{author}{\bibfnamefont{A.}~\bibnamefont{Sterdyniak}},
  \bibinfo{author}{\bibfnamefont{N.}~\bibnamefont{Regnault}}, \bibnamefont{and}
  \bibinfo{author}{\bibfnamefont{B.}~\bibnamefont{Bernevig}},
  \bibinfo{journal}{Physical Review Letters} \textbf{\bibinfo{volume}{106}},
  \bibinfo{pages}{100405} (\bibinfo{year}{2011}).

\bibitem[{\citenamefont{Read and Rezayi}(1999)}]{Read99:RR}
\bibinfo{author}{\bibfnamefont{N.}~\bibnamefont{Read}} \bibnamefont{and}
  \bibinfo{author}{\bibfnamefont{E.}~\bibnamefont{Rezayi}},
  \bibinfo{journal}{Physical Review B} \textbf{\bibinfo{volume}{59}},
  \bibinfo{pages}{8084} (\bibinfo{year}{1999}).

\bibitem[{\citenamefont{Laughlin}(1983)}]{Laughlin83:Nobel}
\bibinfo{author}{\bibfnamefont{R.~B.} \bibnamefont{Laughlin}},
  \bibinfo{journal}{Physical Review Letters} \textbf{\bibinfo{volume}{50}},
  \bibinfo{pages}{1395} (\bibinfo{year}{1983}).

\bibitem[{\citenamefont{Chandran et~al.}(2011)\citenamefont{Chandran, Hermanns,
  Regnault, and Bernevig}}]{Chandran11:ES}
\bibinfo{author}{\bibfnamefont{A.}~\bibnamefont{Chandran}},
  \bibinfo{author}{\bibfnamefont{M.}~\bibnamefont{Hermanns}},
  \bibinfo{author}{\bibfnamefont{N.}~\bibnamefont{Regnault}}, \bibnamefont{and}
  \bibinfo{author}{\bibfnamefont{B.~A.} \bibnamefont{Bernevig}},
  \bibinfo{journal}{ArXiv e-prints}  (\bibinfo{year}{2011}),
  \eprint{1102.2218}.

\bibitem[{\citenamefont{Zozulya et~al.}(2007)\citenamefont{Zozulya, Haque,
  Schoutens, and Rezayi}}]{Zozulya07}
\bibinfo{author}{\bibfnamefont{O.}~\bibnamefont{Zozulya}},
  \bibinfo{author}{\bibfnamefont{M.}~\bibnamefont{Haque}},
  \bibinfo{author}{\bibfnamefont{K.}~\bibnamefont{Schoutens}},
  \bibnamefont{and} \bibinfo{author}{\bibfnamefont{E.}~\bibnamefont{Rezayi}},
  \bibinfo{journal}{Physical Review B} \textbf{\bibinfo{volume}{76}},
  \bibinfo{pages}{125310} (\bibinfo{year}{2007}).

\bibitem[{\citenamefont{Haque et~al.}(2007)\citenamefont{Haque, Zozulya, and
  Schoutens}}]{Haque07:PEE}
\bibinfo{author}{\bibfnamefont{M.}~\bibnamefont{Haque}},
  \bibinfo{author}{\bibfnamefont{O.}~\bibnamefont{Zozulya}}, \bibnamefont{and}
  \bibinfo{author}{\bibfnamefont{K.}~\bibnamefont{Schoutens}},
  \bibinfo{journal}{Physical Review Letters} \textbf{\bibinfo{volume}{98}},
  \bibinfo{pages}{060401} (\bibinfo{year}{2007}).

\bibitem[{\citenamefont{Haque et~al.}(2009)\citenamefont{Haque, Zozulya, and
  Schoutens}}]{Haque09}
\bibinfo{author}{\bibfnamefont{M.}~\bibnamefont{Haque}},
  \bibinfo{author}{\bibfnamefont{O.~S.} \bibnamefont{Zozulya}},
  \bibnamefont{and}
  \bibinfo{author}{\bibfnamefont{K.}~\bibnamefont{Schoutens}},
  \bibinfo{journal}{Journal of Physics A: Mathematical and Theoretical}
  \textbf{\bibinfo{volume}{42}}, \bibinfo{pages}{504012}
  (\bibinfo{year}{2009}).

\bibitem[{\citenamefont{Goerbig}(2011)}]{Goerbig11}
\bibinfo{author}{\bibfnamefont{M.~O.} \bibnamefont{Goerbig}},
  \bibinfo{journal}{ArXiv e-prints}  (\bibinfo{year}{2011}),
  \eprint{1107.1986}.

\bibitem[{\citenamefont{Podolsky and Avron}()}]{Podolsky11:Unpublished}
\bibinfo{author}{\bibfnamefont{D.}~\bibnamefont{Podolsky}} \bibnamefont{and}
  \bibinfo{author}{\bibfnamefont{J.}~\bibnamefont{Avron}},
  \bibinfo{note}{private communication}.

\bibitem[{\citenamefont{Greiter et~al.}(1991)\citenamefont{Greiter, Wen, and
  Wilczek}}]{Greiter91:MR}
\bibinfo{author}{\bibfnamefont{M.}~\bibnamefont{Greiter}},
  \bibinfo{author}{\bibfnamefont{X.-G.} \bibnamefont{Wen}}, \bibnamefont{and}
  \bibinfo{author}{\bibfnamefont{F.}~\bibnamefont{Wilczek}},
  \bibinfo{journal}{Physical Review Letters} \textbf{\bibinfo{volume}{66}},
  \bibinfo{pages}{3205} (\bibinfo{year}{1991}).

\bibitem[{\citenamefont{Simon et~al.}(2007)\citenamefont{Simon, Rezayi, Cooper,
  and Berdnikov}}]{Simon07:Gaffnian}
\bibinfo{author}{\bibfnamefont{S.}~\bibnamefont{Simon}},
  \bibinfo{author}{\bibfnamefont{E.}~\bibnamefont{Rezayi}},
  \bibinfo{author}{\bibfnamefont{N.}~\bibnamefont{Cooper}}, \bibnamefont{and}
  \bibinfo{author}{\bibfnamefont{I.}~\bibnamefont{Berdnikov}},
  \bibinfo{journal}{Physical Review B} \textbf{\bibinfo{volume}{75}},
  \bibinfo{pages}{075317} (\bibinfo{year}{2007}).

\bibitem[{\citenamefont{Sun et~al.}(2011)\citenamefont{Sun, Gu, Katsura, and
  Das~Sarma}}]{Sun11:Flatband}
\bibinfo{author}{\bibfnamefont{K.}~\bibnamefont{Sun}},
  \bibinfo{author}{\bibfnamefont{Z.}~\bibnamefont{Gu}},
  \bibinfo{author}{\bibfnamefont{H.}~\bibnamefont{Katsura}}, \bibnamefont{and}
  \bibinfo{author}{\bibfnamefont{S.}~\bibnamefont{Das~Sarma}},
  \bibinfo{journal}{Physical Review Letters} \textbf{\bibinfo{volume}{106}},
  \bibinfo{pages}{236803} (\bibinfo{year}{2011}).

\bibitem[{\citenamefont{Wang and Ran}(2011)}]{Wang11:Dice}
\bibinfo{author}{\bibfnamefont{F.}~\bibnamefont{Wang}} \bibnamefont{and}
  \bibinfo{author}{\bibfnamefont{Y.}~\bibnamefont{Ran}},
  \bibinfo{journal}{ArXiv e-prints}  (\bibinfo{year}{2011}),
  \eprint{1109.3435}.

\end{thebibliography}
\end{document}